\documentclass{article}

\PassOptionsToPackage{numbers, compress}{natbib}
\usepackage[preprint]{neurips_2023}
\makeatletter
\newlength{\@neuripstablecontentskip}
\renewenvironment{table}
  {\setlength{\abovecaptionskip}{\@neuripsbelowcaptionskip}%
   \setlength{\belowcaptionskip}{\@neuripstablecontentskip}%
   \@float{table}}
  {\end@float}
\makeatother

\usepackage[utf8]{inputenc}
\usepackage[T1]{fontenc}
\usepackage{subcaption}
\usepackage{amssymb}
\usepackage{amsmath}
\usepackage{amsfonts}
\usepackage{url}
\usepackage{booktabs}
\usepackage{longtable}
\usepackage{nicefrac}
\usepackage{xcolor}
\usepackage{graphicx}
\usepackage{tikz}
\usetikzlibrary{
  arrows.meta,
  positioning,
  fit,
  calc,
  backgrounds,
  shapes.geometric
}
\providecommand{\omniStyleLoaded}{0}
\ifnum\omniStyleLoaded=0
\def\omniStyleLoaded{1}

\definecolor{omniBlue}{RGB}{10,50,110}
\definecolor{omniBlueSoft}{RGB}{70,110,170}
\definecolor{omniBlueFill}{RGB}{232,240,250}
\definecolor{omniCard}{RGB}{255,255,255}
\definecolor{omniTeal}{RGB}{56,151,152}
\definecolor{omniTealDark}{RGB}{40,120,122}
\definecolor{omniTealFill}{RGB}{230,245,245}
\definecolor{omniInk}{RGB}{20,30,50}
\definecolor{omniGray}{RGB}{120,130,145}
\definecolor{omniGrayFill}{RGB}{242,244,247}
\definecolor{omniOrange}{RGB}{180,110,40}
\definecolor{omniOrangeFill}{RGB}{255,245,230}

\tikzset{
  every node/.append style={outer sep=0pt},
  omni arrow/.style={
    -{Stealth[length=2.2mm, width=1.6mm]},
    omniBlue,
    line width=0.8pt,
  },
  omni ctrl/.style={
    -{Stealth[length=2.2mm, width=1.6mm]},
    omniInk,
    line width=0.8pt,
  },
  omni heavy/.style={
    -{Stealth[length=2.2mm, width=1.6mm]},
    omniBlueSoft,
    dashed,
    dash pattern=on 2.5pt off 2pt,
    line width=0.8pt,
  },
  omni heavy bi/.style={
    {Stealth[length=2.0mm, width=1.5mm]}-{Stealth[length=2.0mm, width=1.5mm]},
    omniBlueSoft,
    dashed,
    dash pattern=on 2.5pt off 2pt,
    line width=0.8pt,
  },
  omni layer/.style={
    draw=omniBlue,
    fill=omniBlueFill,
    line width=0.7pt,
    rounded corners=3pt,
    outer sep=0pt,
  },
  omni teal layer/.style={
    draw=omniTealDark,
    fill=omniTealFill,
    line width=0.7pt,
    rounded corners=3pt,
    outer sep=0pt,
  },
  omni card/.style={
    draw=omniBlueSoft,
    fill=omniCard,
    line width=0.6pt,
    rounded corners=2pt,
    align=center,
    font=\sffamily\footnotesize,
    outer sep=0pt,
  },
  omni teal card/.style={
    draw=omniTeal,
    fill=omniCard,
    line width=0.6pt,
    rounded corners=2pt,
    align=center,
    font=\sffamily\footnotesize,
    outer sep=0pt,
  },
  omni gray card/.style={
    draw=omniGray,
    fill=omniGrayFill,
    line width=0.6pt,
    rounded corners=2pt,
    align=center,
    font=\sffamily\footnotesize,
    outer sep=0pt,
  },
  omni session/.style={
    draw=omniOrange,
    fill=omniOrangeFill,
    line width=0.7pt,
    rounded corners=3pt,
    align=left,
    font=\sffamily\footnotesize,
    outer sep=0pt,
  },
  omni layer title/.style={
    font=\sffamily\bfseries\small,
    text=omniBlue,
    outer sep=0pt,
  },
  omni fig title/.style={
    font=\sffamily\bfseries\small,
    text=omniInk,
    inner sep=0pt,
    outer sep=0pt,
  },
  omni runtime/.style={
    draw=none,
    fill=omniTeal,
    rounded corners=3pt,
    outer sep=0pt,
  },
  omni runtime title/.style={
    font=\sffamily\bfseries\small,
    text=white,
    outer sep=0pt,
  },
  omni runtime card/.style={
    draw=none,
    fill=omniCard,
    rounded corners=2pt,
    align=center,
    font=\sffamily\bfseries\footnotesize,
    text=omniTealDark,
    minimum height=7.5mm,
    outer sep=0pt,
  },
  omni scope line/.style={
    omniBlue,
    dashed,
    dash pattern=on 2.2pt off 2.2pt,
    line width=0.7pt,
  },
  omni process/.style={
    draw=omniBlueSoft,
    fill=omniBlueFill,
    line width=0.6pt,
    rounded corners=2pt,
    align=center,
    font=\sffamily\footnotesize,
    minimum height=8mm,
    outer sep=0pt,
  },
  omni process teal/.style={
    draw=omniTeal,
    fill=omniTealFill,
    line width=0.6pt,
    rounded corners=2pt,
    align=center,
    font=\sffamily\footnotesize,
    minimum height=8mm,
    outer sep=0pt,
  },
  omni decision/.style={
    diamond,
    aspect=2.2,
    draw=omniTeal,
    fill=omniTealFill,
    line width=0.6pt,
    align=center,
    font=\sffamily\footnotesize,
    inner sep=1pt,
    outer sep=0pt,
  },
  omni step/.style={
    circle,
    fill=omniBlue,
    text=white,
    font=\sffamily\bfseries\scriptsize,
    inner sep=1pt,
    minimum size=4.2mm,
    outer sep=0pt,
  },
  omni step teal/.style={
    circle,
    fill=omniTealDark,
    text=white,
    font=\sffamily\bfseries\scriptsize,
    inner sep=1pt,
    minimum size=4.2mm,
    outer sep=0pt,
  },
  omni connector box/.style={
    draw=omniBlueSoft,
    dashed,
    dash pattern=on 2.5pt off 2pt,
    fill=omniCard,
    line width=0.7pt,
    rounded corners=3pt,
    align=center,
    font=\sffamily\bfseries\footnotesize,
    text=omniBlue,
    outer sep=0pt,
  },
}

\newcommand{\omniBalanceLR}{%
  \path ([xshift=-3.2pt]current bounding box.west);%
}

\fi

\usepackage{placeins}
\usepackage{float}
\usepackage{multirow}
\usepackage{makecell}
\usepackage{enumitem}
\usepackage{setspace}
\usepackage{xspace}
\usepackage{ragged2e}
\usepackage{fancyhdr}
\usepackage[most]{tcolorbox}
\usepackage{titlesec}
\usepackage{natbib}
\usepackage[colorlinks=true, linkcolor=blue, urlcolor=blue, citecolor=blue]{hyperref}
\usepackage{cleveref}

\newcommand{\name}{vLLM-Omni}
\newcommand{\sys}{vLLM-Omni\xspace}
\newcommand{\stitle}[1]{\noindent\textbf{#1.} }

\definecolor{vllmBlue}{RGB}{48,160,248}

\titleformat{\section}
  {\Large\bfseries}{\thesection}{1em}{}
\titleformat{\subsection}
  {\large\bfseries}{\thesubsection}{1em}{}
\titleformat{\subsubsection}
  {\normalsize\bfseries}{\thesubsubsection}{1em}{}
\makeatletter
\let\omni@orig@section\section
\renewcommand{\section}{\FloatBarrier\omni@orig@section}
\makeatother

\begin{document}

\pagestyle{fancy}
\fancyhf{}
\setlength{\headwidth}{\textwidth}
\renewcommand{\headrulewidth}{0.4pt}
\chead{vLLM-Omni Technical Report}
\cfoot{\thepage}

\fancypagestyle{titlepage}{
  \fancyhf{}
  \setlength{\headwidth}{\textwidth}
  \renewcommand{\headrulewidth}{0pt}
  \lfoot{\sffamily\small 2026 Q3}
  \cfoot{\thepage}
}

\thispagestyle{titlepage}
  \vspace*{-0.85em}%
  \noindent\includegraphics[height=1.15cm]{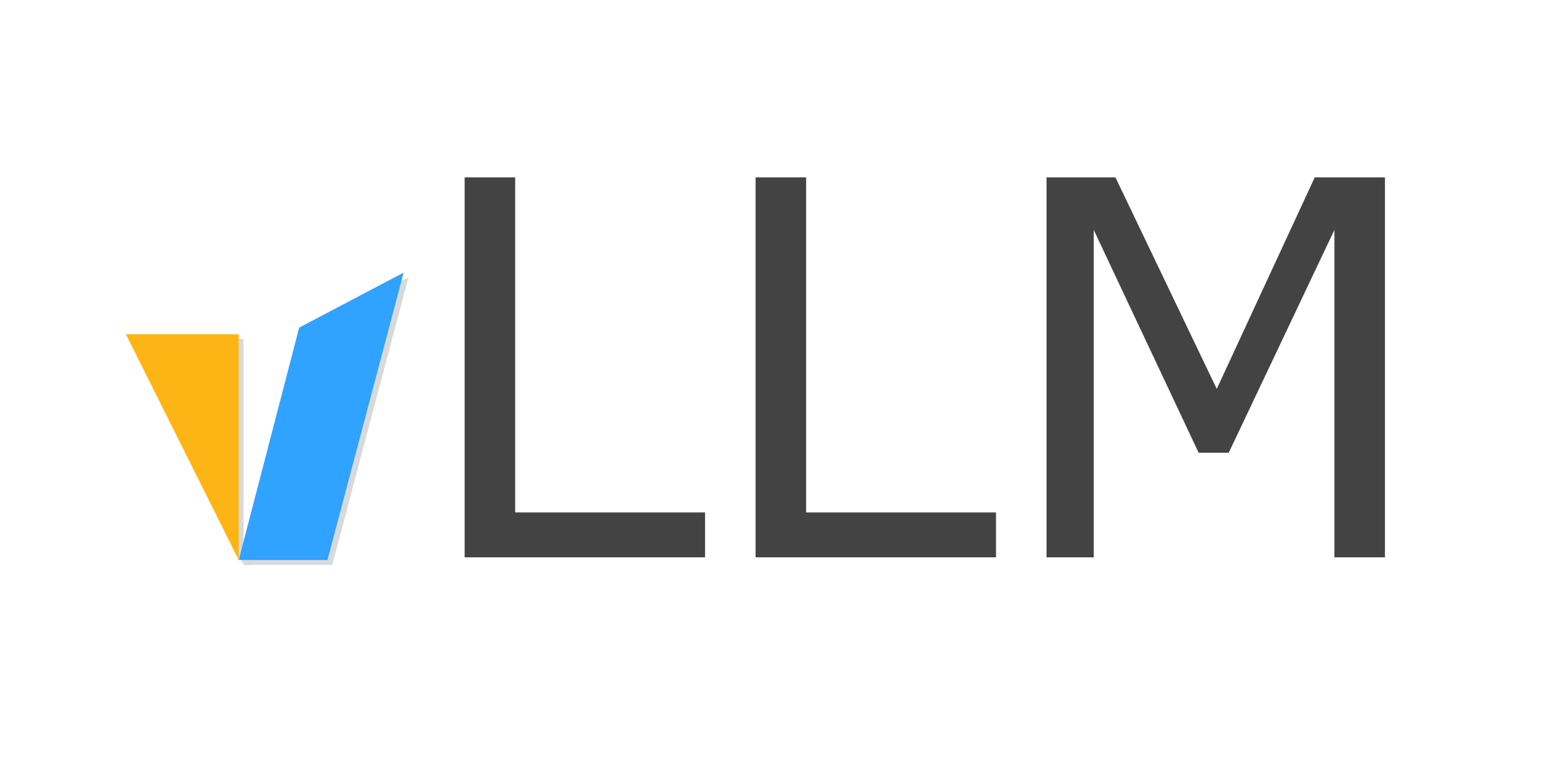}\par
  \vspace{0.55em}%
  {\centering
    {\color{vllmBlue}\rule{\linewidth}{1.1pt}}\par
    \nointerlineskip\vspace{0.65em}%
    {\sffamily\Large\bfseries\color{omniInk}%
      vLLM-Omni Technical Report:}\par
    \vspace{0.28em}%
    {\sffamily\large\bfseries\color{omniInk}%
      A Unified Serving Runtime for Omni-Modality Generation}\par
    \vspace{0.65em}%
    \nointerlineskip
    {\color{vllmBlue}\rule{\linewidth}{1.1pt}}\par
    \vspace{0.85em}%
    {\sffamily\large\textit{vLLM-Omni Team}}\par
    \vspace{0.55em}%
    {\sffamily\small\href{https://github.com/vllm-project/vllm-omni}{%
      \textcolor{vllmBlue}{\texttt{github.com/vllm-project/vllm-omni}}}}\par
    \vspace{1.3em}%
  }%
  \begin{tcolorbox}[
    enhanced,
    colback=white,
    colframe=vllmBlue,
    boxrule=0.55pt,
    arc=2.5mm,
    left=7mm,
    right=7mm,
    top=4.5mm,
    bottom=4.5mm,
    before skip=0pt,
    after skip=6pt,
  ]
    {\centering
      \sffamily\large\bfseries\color{vllmBlue}Abstract\par}
    \vspace{2.2mm}
    \justifying
    \setlength{\parindent}{0pt}%
    \setlength{\parskip}{0.45em}%
    Interaction with intelligent systems is rapidly expanding beyond text-centric
chatbots and coding agents.  Speech-native assistants, visual generation and
editing, world-model environments, and robot action loops require models that
emit text, audio, images, video, and actions.
These models further differ in execution pattern: multi-stage autoregressive
omni and TTS pipelines, iterative diffusion or flow-matching generators, and
longer-lived world-model or robot loops that carry state across steps.  As a
result, serving is no longer a single text decode loop, but a heterogeneous
multi-stage workflow with cross-stage transfer, streaming, and session-shaped
interaction.

Existing inference stacks are typically optimized for one architecture family
at a time.  LLM servers deepen autoregressive scheduling and KV management,
while diffusion stacks deepen denoising and parallel generation.  Neither
provides a shared control plane for pipelines that emit speech, pixels, or
actions through separate generators, so production deployments often fall back
to ad-hoc composition across disjoint runtimes.

We present \textbf{vLLM-Omni}, a unified serving runtime for omni-modality
generation.  \sys{} organizes each workload as a multi-stage pipeline under a
single orchestrator that admits requests, advances them across stages, and
demultiplexes streaming outputs.  Specialized engines and stage replicas
provide compute; a connector carries heavy payloads on the data plane; and
session-oriented control supports long-lived duplex, world-model, and robot
workloads on the same runtime.

This technical report describes the architecture of \sys{}: stage-level KV
paths, replica pools, multi-hardware platforms, and an efficiency stack on top
of the original stage-based design.  OpenAI-compatible and OpenPI surfaces map
these mechanisms onto omni, TTS, image and video generation, world-model,
robot, and full-duplex workloads.  We evaluate the system against the public
multimodal nightly CI on NVIDIA H100, using frozen nightly recipe
results; the TTS and MiniCPM-o tables are measured separately on H200 to compare
default deployments with the fused single-stage Qwen3-TTS and MRv2 profiles. The campaign centers on Qwen3-Omni speech serving with async-chunk and
replica sweeps, and adds TTS, image and video diffusion, Cosmos3/MiniMax video
cells, and MiniCPM-o duplex health metrics.

  \end{tcolorbox}%

\clearpage

\tableofcontents
\clearpage

\section{Introduction}
\label{sec:intro}

\begin{figure}[t]
  \centering
  \resizebox{\linewidth}{!}{
\begin{tikzpicture}[
  x=1cm, y=1cm,
  font=\sffamily,
]
  \def\layerW{14.2}
  \def\cardW{3.05}
  \def\cardH{1.60}
  \def\wlH{1.40}
  \def\gap{0.35}
  \pgfmathsetmacro{\xOne}{0.55 + 0.5*\cardW}
  \pgfmathsetmacro{\xTwo}{\xOne + \cardW + \gap}
  \pgfmathsetmacro{\xThree}{\xTwo + \cardW + \gap}
  \pgfmathsetmacro{\xFour}{\xThree + \cardW + \gap}
  \pgfmathsetmacro{\layerCX}{0.5*\layerW}

  \node[omni layer, minimum width=\layerW cm, minimum height=2.60cm]
    (app) at (\layerCX, 6.15) {};
  \node[omni layer title, anchor=north] at ($(app.north)+(0,-0.12)$)
    {Application Scenarios};

  \node[omni card, minimum width=\cardW cm, minimum height=\cardH cm]
    (sc1) at (\xOne, 5.85) {};
  \node[omni card, minimum width=\cardW cm, minimum height=\cardH cm]
    (sc2) at (\xTwo, 5.85) {};
  \node[omni card, minimum width=\cardW cm, minimum height=\cardH cm]
    (sc3) at (\xThree, 5.85) {};
  \node[omni card, minimum width=\cardW cm, minimum height=\cardH cm]
    (sc4) at (\xFour, 5.85) {};

  \begin{scope}[shift={(\xOne, 6.15)}, omniBlue]
    \draw[line width=0.7pt] (0,0) circle[radius=0.30];
    \foreach \dx/\h in {-0.14/0.10, -0.05/0.18, 0.05/0.14, 0.14/0.20} {
      \draw[line width=0.75pt, line cap=round] (\dx,-\h) -- (\dx,\h);
    }
  \end{scope}
  \begin{scope}[shift={(\xTwo, 6.15)}, omniBlue]
    \fill[omniBlueFill, rounded corners=1.1pt]
      (-0.34,-0.16) rectangle (0.00,0.16);
    \draw[line width=0.75pt, rounded corners=1.1pt]
      (-0.34,-0.16) rectangle (0.00,0.16);
    \fill[omniBlueSoft] (-0.24,0.06) circle[radius=0.028];
    \draw[line width=0.6pt]
      (-0.34,-0.06) .. controls (-0.26,0.05) and (-0.18,0.04) .. (-0.12,-0.02)
      .. controls (-0.06,0.07) and (-0.02,0.05) .. (0.00,-0.03);
    \fill[omniCard, rounded corners=1.1pt]
      (0.06,-0.16) rectangle (0.34,0.16);
    \draw[line width=0.75pt, rounded corners=1.1pt]
      (0.06,-0.16) rectangle (0.34,0.16);
    \fill[omniBlue]
      (0.13,-0.08) -- (0.13,0.08) -- (0.29,0.00) -- cycle;
  \end{scope}

  \begin{scope}[shift={(\xThree, 6.15)}, omniBlue]
    \draw[line width=0.7pt] (0,0) circle[radius=0.30];
    \draw[line width=0.55pt] (0,-0.30) -- (0,0.30);
    \draw[line width=0.55pt] (-0.30,0) -- (0.30,0);
    \draw[line width=0.55pt]
      (-0.12,-0.27) .. controls (-0.20,-0.08) and (-0.20,0.08) .. (-0.12,0.27);
    \draw[line width=0.55pt]
      (0.12,-0.27) .. controls (0.20,-0.08) and (0.20,0.08) .. (0.12,0.27);
  \end{scope}
  \begin{scope}[shift={(\xFour, 6.12)}, omniBlue]
    \fill (-0.22,-0.26) rectangle (-0.06,-0.18);
    \draw[line width=0.75pt, line cap=round, line join=round]
      (-0.14,-0.18) -- (-0.14,0.00) -- (0.06,0.14) -- (0.24,0.00);
    \draw[line width=0.7pt, line cap=round]
      (0.24,0.00) -- (0.34,0.10);
    \draw[line width=0.7pt, line cap=round]
      (0.24,0.00) -- (0.34,-0.08);
  \end{scope}

  \node[font=\sffamily\bfseries\footnotesize, text=omniBlue, anchor=north]
    at (\xOne, 5.70) {Speech Assistants};
  \node[font=\sffamily\bfseries\footnotesize, text=omniBlue, anchor=north]
    at (\xTwo, 5.70) {Image / Video Gen.};
  \node[font=\sffamily\bfseries\footnotesize, text=omniBlue, anchor=north]
    at (\xThree, 5.70) {World Models};
  \node[font=\sffamily\bfseries\footnotesize, text=omniBlue, anchor=north]
    at (\xFour, 5.70) {Robot Action};

  \foreach \x in {\xOne,\xTwo,\xThree,\xFour} {
    \draw[omni arrow] (\x, 4.85) -- (\x, 4.45);
  }

  \node[omni layer, minimum width=\layerW cm, minimum height=2.40cm]
    (wl) at (\layerCX, 3.25) {};
  \node[omni layer title, anchor=north] at ($(wl.north)+(0,-0.12)$)
    {Model Workloads};

  \node[omni card, minimum width=\cardW cm, minimum height=\wlH cm]
    (wl1) at (\xOne, 2.95) {};
  \node[omni card, minimum width=\cardW cm, minimum height=\wlH cm]
    (wl2) at (\xTwo, 2.95) {};
  \node[omni card, minimum width=\cardW cm, minimum height=\wlH cm]
    (wl3) at (\xThree, 2.95) {};
  \node[omni card, minimum width=\cardW cm, minimum height=\wlH cm]
    (wl4) at (\xFour, 2.95) {};

  \node[font=\sffamily\bfseries\footnotesize, text=omniBlue, align=center]
    at (\xOne, 3.15) {AR Omni / TTS};
  \node[font=\sffamily\scriptsize, text=omniBlue, align=center]
    at (\xOne, 2.78) {thinker $\rightarrow$ talker $\rightarrow$ wav};

  \node[font=\sffamily\bfseries\footnotesize, text=omniBlue, align=center]
    at (\xTwo, 3.15) {DiT / Hybrid};
  \node[font=\sffamily\scriptsize, text=omniBlue, align=center]
    at (\xTwo, 2.78) {(AR $\rightarrow$) DiT denoise};

  \node[font=\sffamily\bfseries\footnotesize, text=omniBlue, align=center]
    at (\xThree, 3.15) {World Session};
  \node[font=\sffamily\scriptsize, text=omniBlue, align=center]
    at (\xThree, 2.78) {predict $\rightarrow$ recondition};

  \node[font=\sffamily\bfseries\footnotesize, text=omniBlue, align=center]
    at (\xFour, 3.15) {VLA / Policy};
  \node[font=\sffamily\scriptsize, text=omniBlue, align=center]
    at (\xFour, 2.78) {obs $\rightarrow$ action chunks};

  \node[omni runtime, minimum width=\layerW cm, minimum height=1.90cm]
    (rt) at (\layerCX, 0.55) {};
  \node[omni runtime title] at (\layerCX, 1.28) {vLLM-Omni Unified Runtime};

  \node[omni runtime card, minimum width=4.0cm]
    (rt1) at (2.35, 0.32) {Stage Graph};
  \node[omni runtime card, minimum width=4.6cm]
    (rt2) at (\layerCX, 0.32) {OpenAI-compatible APIs};
  \node[omni runtime card, minimum width=3.2cm]
    (rt3) at (11.85, 0.32) {OpenPI};

  \draw[white, line width=0.7pt]
    ($(rt1.east)!0.5!(rt2.west)$) ++(0,-0.28) -- ++(0,0.56);
  \draw[white, line width=0.7pt]
    ($(rt2.east)!0.5!(rt3.west)$) ++(0,-0.28) -- ++(0,0.56);

  \foreach \x in {\xOne,\xTwo,\xThree,\xFour} {
    \draw[omni arrow] (\x, 2.05) -- (\layerCX, 1.50);
  }

  \draw[omni scope line] (-0.35, 7.50) -- (-0.35, -0.40);
  \node[font=\sffamily\bfseries\small, text=omniBlue, rotate=90]
    at (-0.70, 3.55) {Serving scope};

  \omniBalanceLR
\end{tikzpicture}}
  \caption{Scope map from application scenarios to the multi-stage model families and serving surfaces considered by \sys{}.}
  \label{fig:intro-scope}
\end{figure}
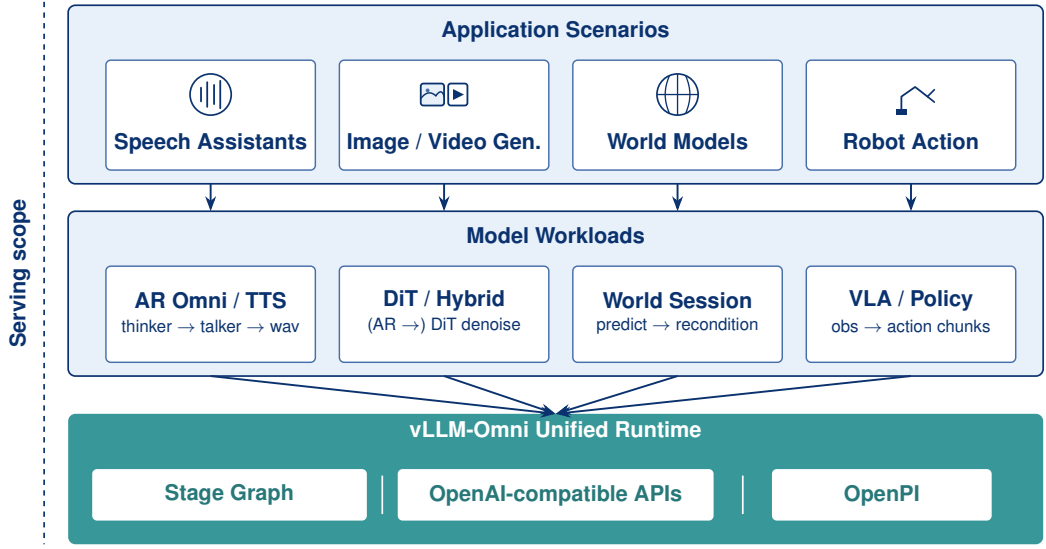

Generative-model serving now spans several execution patterns beyond
autoregressive text decoding~\cite{gpt4,gemini,qwen3}.
Omni-modality models take mixed text, audio, image, and video inputs and produce text and speech within one model family~\cite{Qwen2.5-Omni,Qwen3-Omni,Ming-Omni}.
Dedicated text-to-speech (TTS) models focus on high-quality and often streaming speech synthesis from text or LLM outputs~\cite{cosyvoice,cosyvoice2,cosyvoice3}.
Visual generation models synthesize or edit images and videos with diffusion-style or hybrid generators~\cite{Qwen-Image,wan,FLUX.1}.
World foundation models predict future world states as video conditioned on past observations~\cite{cosmos}.
Vision--language--action (VLA) models map perception and language to robot actions~\cite{openvla,grootn1}.
Figure~\ref{fig:intro-scope} maps these model families to the serving scope
considered in this report.

These model families differ not only in modality, but also in execution pattern.
Omni models often run a multi-stage autoregressive (AR) pipeline in which a thinker stage generates text or intermediate representations, a talker stage emits speech codecs, and a decoder reconstructs waveforms~\cite{Qwen2.5-Omni,Qwen3-Omni}.
TTS systems typically couple an AR or LLM-based codec predictor with a flow-matching or vocoder stage, and they are frequently served with chunked streaming for low first-packet latency~\cite{cosyvoice2,cosyvoice3}.
Image and video generators usually iterate a diffusion transformer (DiT) or flow-matching denoiser over many steps, sometimes after an AR or multimodal understanding front-end~\cite{DiT,Qwen-Image,wan}.
World models generate future video from past frames and text or action conditions; in long rollouts, previously predicted frames become the conditioning context for later steps~\cite{cosmos}.
VLA policies repeatedly map observations and instructions to action chunks, often through a vision--language backbone followed by an action head or diffusion action module~\cite{openvla,grootn1}.
From a systems perspective, these workloads no longer resemble a single text decode loop.
Execution may span heterogeneous stages with different batching and memory behavior, while input--output connectivity is also more diverse: a world-model step may feed predicted video back as the next observation, a duplex speech agent may keep accepting audio while producing speech, and a robot loop may stream actions against a continuously updating sensory context.

This diversity creates a serving problem that single-paradigm inference stacks do not cover cleanly.
LLM serving systems such as vLLM and SGLang excel at autoregressive decoding, KV-cache management, and continuous batching for text~\cite{vllm,sglang}, and in many cases they also support multimodal input.
They are not, however, designed as a native runtime for multi-stage pipelines that emit speech, pixels, or actions through separate generators.
Diffusion-oriented stacks optimize denoising, caching, and parallelism for DiT workloads~\cite{diffusers,fang2024xdit}, yet they do not provide a shared control plane for AR stages, cross-stage transfer, or session-oriented interaction.
Deploying these components as disjoint services duplicates admission and
lifecycle logic and prevents stage-level streaming and KV transfer from being
managed as one request.

The original \sys{} system introduced a stage graph, disaggregated engines,
and connector-backed transfer for any-to-any models~\cite{vllmomni}.
SGLang-Omni and M* independently explore stage-oriented and graph-walk
execution for composite models~\cite{sglangomni,mstar2026}.
This report describes the architecture of \sys{}: it retains the stage
abstraction while making orchestration, streaming progress, replica pools,
intermediate-data movement, and sessions explicit runtime concerns.
A stage owns one model component, one execution engine, and a resource budget; the orchestrator owns request admission, cross-stage advancement, and client-visible progress.
This split allows omni, TTS, diffusion, world-model, and robot workloads to share the same runtime machinery while still specializing compute inside each stage.

Around that control plane, \sys{} adds the modules needed for online
multimodal serving: specialized engines and stage replicas, intermediate-data
management for hidden states, codec codes, latents, and KV blocks, a connector
transport for those payloads, and session-oriented control for long-lived
duplex, world-model, and robot loops.
Streaming input and output are treated as request progress on the orchestrator; sessions extend the same machinery when interaction outlives a single generate.
Section~\ref{sec:architecture} develops these pieces in detail.

To summarize, this report makes the following contributions:
\begin{enumerate}[leftmargin=*]
\item We present the system architecture of \sys{} as a multi-stage serving framework centered on the orchestrator, covering execution engines, intermediate-data management, the connector, and session-oriented control.
\item We describe platform and ecosystem support, including multi-hardware backends, the efficiency stack, and training--serving integration such as RL handoff.
\item We map OpenAI-compatible and OpenPI serving interfaces onto supported model families across omni, TTS, diffusion, world-model, VLA, and full-duplex workloads.
\item We report a systems evaluation drawn from the public multimodal nightly CI on NVIDIA H100 (with the TTS and MiniCPM-o tables measured on H200), centered on Qwen3-Omni speech serving with async-chunk and replica sweeps, together with TTS, image/video diffusion, Cosmos3/MiniMax video cells, and MiniCPM-o duplex health metrics.
\end{enumerate}

\section{vLLM-Omni System Architecture}
\label{sec:architecture}

\subsection{Overview}
\label{sec:arch-overview}

\begin{figure}[t]
  \centering
  \resizebox{\linewidth}{!}{
\begin{tikzpicture}[
  x=1cm, y=1cm,
  font=\sffamily,
]
  \def\bandW{11.0}
  \def\g{0.55} 
  \def\hEntry{2.05}
  \def\hOrch{2.05}
  \def\hPipe{2.05}
  \def\hConn{0.58}
  \pgfmathsetmacro{\yEntry}{7.15}
  \pgfmathsetmacro{\yOrch}{\yEntry - 0.5*\hEntry - \g - 0.5*\hOrch}
  \pgfmathsetmacro{\yPipe}{\yOrch - 0.5*\hOrch - \g - 0.5*\hPipe}
  \pgfmathsetmacro{\yConn}{\yPipe - 0.5*\hPipe - \g - 0.5*\hConn}
  \pgfmathsetmacro{\yApi}{\yEntry + 0.5*\hEntry + 0.55}

  \node[omni gray card, minimum width=4.0cm, minimum height=0.58cm]
    (api1) at (3.20, \yApi) {OpenAI / WebSocket};
  \node[omni gray card, minimum width=4.0cm, minimum height=0.58cm]
    (api2) at (7.80, \yApi) {OpenPI / Omni API};

  \node[omni layer, minimum width=\bandW cm, minimum height=\hEntry cm]
    (entry) at (5.5, \yEntry) {};
  \node[omni layer title, anchor=north west]
    at ($(entry.north west)+(0.18,-0.10)$) {Entrypoint};
  \node[omni card, fill=omniBlueFill, minimum width=4.0cm, minimum height=1.15cm]
    (async) at (3.20, \yEntry-0.22) {};
  \node[font=\sffamily\bfseries\footnotesize, text=omniBlue]
    at (3.20, \yEntry+0.08) {AsyncOmni};
  \node[font=\sffamily\scriptsize, text=omniInk, align=center]
    at (3.20, \yEntry-0.38) {\texttt{request\_queue}\\\texttt{output\_queue}};
  \node[omni card, minimum width=4.0cm, minimum height=1.15cm]
    (aeng) at (7.80, \yEntry-0.22) {};
  \node[font=\sffamily\bfseries\footnotesize, text=omniBlue]
    at (7.80, \yEntry+0.08) {AsyncOmni Engine};
  \node[font=\sffamily\scriptsize, text=omniInk, align=center]
    at (7.80, \yEntry-0.38) {admit / abort\\client output demux};
  \draw[omniBlue, line width=0.8pt, -{Stealth[length=2.2mm, width=1.6mm]}] (async.east) -- (aeng.west);

  \draw[omni ctrl] (api1.south) -- (async.north);
  \draw[omni ctrl] (api2.south) -- (aeng.north);

  \node[omni layer, minimum width=\bandW cm, minimum height=\hOrch cm]
    (orch) at (5.5, \yOrch) {};
  \node[omni layer title, anchor=north west]
    at ($(orch.north west)+(0.18,-0.10)$) {Control plane: Orchestrator};
  \node[omni card, minimum width=2.15cm, minimum height=0.62cm]
    (lc1) at (1.85, \yOrch+0.15) {\scriptsize\textbf{Admit}};
  \node[omni card, minimum width=2.15cm, minimum height=0.62cm]
    (lc2) at (4.3, \yOrch+0.15) {\scriptsize\textbf{Advance}};
  \node[omni card, minimum width=2.15cm, minimum height=0.62cm]
    (lc3) at (6.7, \yOrch+0.15) {\scriptsize\textbf{Stream}};
  \node[omni card, minimum width=2.15cm, minimum height=0.62cm]
    (lc4) at (9.15, \yOrch+0.15) {\scriptsize\textbf{Complete}};
  \draw[omni arrow] (lc1) -- (lc2);
  \draw[omni arrow] (lc2) -- (lc3);
  \draw[omni arrow] (lc3) -- (lc4);
  \node[omni teal card, minimum width=10.0cm, minimum height=0.48cm]
    (sess) at (5.5, \yOrch-0.55)
    {\scriptsize\textbf{Session:} lifecycle $\cdot$ state continuity $\cdot$ concurrent I/O};

  \draw[omni ctrl] (entry.south) -- (orch.north);

  \node[omni teal layer, minimum width=\bandW cm, minimum height=\hPipe cm]
    (pipe) at (5.5, \yPipe) {};
  \node[font=\sffamily\bfseries\small, text=omniTealDark, anchor=north west]
    at ($(pipe.north west)+(0.18,-0.08)$) {Execution stages};
  \node[omni teal card, minimum width=2.8cm, minimum height=0.62cm]
    (s0) at (2.0, \yPipe-0.18) {\textbf{Stage 0}};
  \node[omni teal card, minimum width=2.8cm, minimum height=0.62cm]
    (s1) at (5.5, \yPipe-0.18) {\textbf{Stage 1}};
  \node[omni teal card, minimum width=2.8cm, minimum height=0.62cm]
    (s2) at (9.0, \yPipe-0.18) {\textbf{Stage 2}};
  \draw[omni arrow] (s0) -- (s1);
  \draw[omni arrow] (s1) -- (s2);
  \foreach \n in {s0,s1,s2} {
    \draw[omni ctrl] (\n.north |- orch.south) -- (\n.north);
  }

  \node[omni connector box, minimum width=\bandW cm, minimum height=\hConn cm]
    (conn) at (5.5, \yConn) {OmniConnector};
  \foreach \n in {s0,s1,s2} {
    \draw[omniInk, line width=0.65pt]
      (\n.south) -- (\n.south |- conn.north);
  }

  \omniBalanceLR
\end{tikzpicture}}
  \caption{System overview of \sys{}: APIs enter through AsyncOmni and its engine facade, the orchestrator advances stages (with optional session control), and OmniConnector carries inter-stage payloads.}
  \label{fig:arch-overview}
\end{figure}
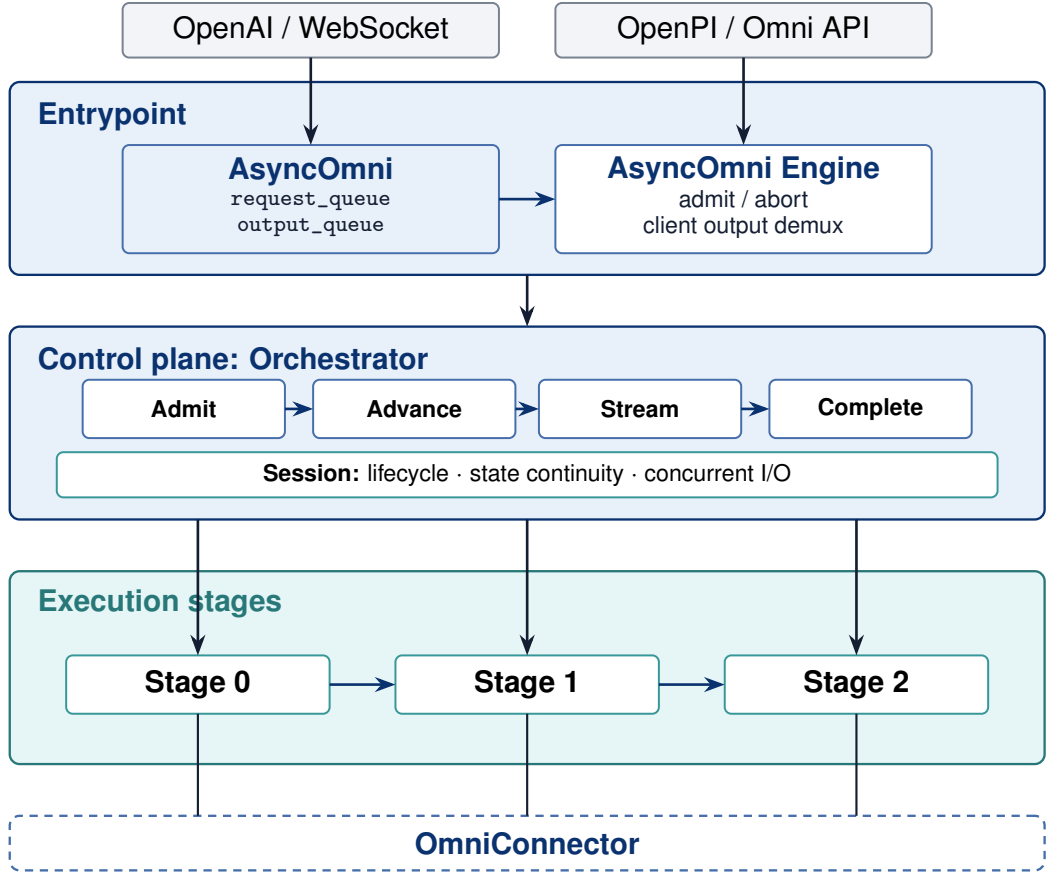

\sys{} is organized as a multi-stage serving framework rather than as a single generation engine with modality plugins.
A stage binds one model component to one execution engine and an explicit resource budget.
Stages exchange typed intermediate state (hidden states, codec codes, latents, packed KV blocks, and related tensors) and may be co-located or disaggregated across processes, devices, and hosts.
Above the stages, a single orchestrator owns request admission, cross-stage advancement, client-visible progress, and completion.
Figure~\ref{fig:arch-overview} places these pieces in one view and marks where the remainder of this section expands them.

The architectural split is intentional.
The orchestrator decides \emph{which} stage runs next, \emph{which} replica receives the work, and \emph{when} a result is client-visible versus a stage handoff.
It does not schedule tokens inside a stage, manage paged KV, denoise diffusion steps, or move large tensors.
Those concerns stay with execution engines, intermediate-data management, and OmniConnector.
This boundary lets omni, TTS, diffusion, world-model, and robot pipelines share one progress machine while specializing compute per stage.

\paragraph{Configuration Composition.}
Deployments are assembled from three layers.
A pipeline registry maps each supported \texttt{model\_type} to a pipeline that declares logical stages, default engines, and bridge processors.
A deploy overlay specializes placement, connector choice, streaming edges, sampling defaults, and hardware overrides without rewriting the pipeline.
An optional parallel strategy expands tensor-parallel, data-parallel, and stage-replica axes into concrete device layouts.
Entrypoints resolve these layers through one factory path, so chat, speech, diffusion, and robot deployments share the same composition rule while keeping model topology separate from cluster placement.

\paragraph{Module Map.}
The rest of this section follows the modules in Figure~\ref{fig:arch-overview}.
Section~\ref{sec:arch-entrypoint-orch} covers entrypoint design and the orchestrator, including streaming as request progress.
Section~\ref{sec:arch-engines} covers execution engines and stage replicas.
Sections~\ref{sec:arch-kv} and~\ref{sec:arch-connector} cover intermediate-data management and the connector transport.
Section~\ref{sec:arch-session} covers session-oriented control as an extension of the same control plane for long-lived interaction, not as a second runtime.

\subsection{Entrypoint Design}
\label{sec:arch-entrypoint-orch}

Multi-stage omni pipelines need one place where requests are admitted, progressed, and observed by clients.
Without that control plane, each API surface tends to reinvent stage routing, partial-output delivery, and cancellation.
\sys{} therefore centers the framework on an orchestrator and separates client-facing I/O from stage execution.
Entrypoints translate external calls into engine requests.
The orchestrator owns request state, logical stage advancement, and the distinction between outputs destined for clients and payloads destined for downstream stages.
Streaming is handled here as part of request progress, not as a parallel serving stack.

\paragraph{Framework Boundary.}
The orchestrator decides which stage should run next for a request, which replica receives that work, and when a client-visible output or a stage handoff is warranted.
It does not schedule tokens inside a stage, manage paged KV, denoise diffusion steps, or copy large tensors between workers.
Those concerns remain with execution engines and with OmniConnector
(Sections~\ref{sec:arch-engines} and~\ref{sec:arch-connector}).
Stage~0 tokenization and multimodal preprocessing also stay off the orchestrator loop: the entrypoint builds a ready engine request before admission, so the control loop admits runnable work rather than parsing prompts.
As a result, new modalities can be added by introducing stages, engines, or connectors without forking a second request-progress machine.

\paragraph{Request Admission.}
\sys{} exposes two programmatic entrypoints that share one engine stack: a synchronous offline path and an asynchronous online path used by production servers.
OpenAI-compatible HTTP and WebSocket services, together with OpenPI robot serving, are thin adapters over the asynchronous path.
They decode protocol messages into engine requests and relay streamed engine outputs to clients.
Protocol schemas are deferred to Section~\ref{sec:apis}.
Architecturally, all surfaces converge on one path:
AsyncOmni and its engine facade (\texttt{request\_queue} / \texttt{output\_queue}, admit/abort, and client-output demultiplexing), the orchestrator, and per-stage pools shown in Figure~\ref{fig:arch-overview}.
This convergence allows chat, speech, video, and robot workloads to reuse the same advancement and streaming machinery.
AsyncOmni also exposes pause/resume and sleep/wake on AR stages so operators can suspend serving without tearing down the orchestrator or reloading pipelines, which complements the train--serve sleep path in Section~\ref{sec:platforms-efficiency}.

\paragraph{Control-Plane State.}
The orchestrator runs on a dedicated control loop and keeps per-request state for every active request.
That state records the pipeline endpoints, per-stage sampling configuration, sticky stage-submit watermarks, and bookkeeping for resumable streaming.
Completion is modality-set completion rather than arrival at a single sink stage.
Each request carries a final-output stage set, and the request finishes only when every stage in that set has finished.
An omni speech request can therefore continue streaming thinker text while waveform reconstruction is still producing audio, and become complete only after both modalities have drained.
Prefill--decode overlays additionally retain transfer identifiers and multimodal side outputs on this state so that decode stages can resume without reconstructing the request through the client API
(Section~\ref{sec:arch-kv}).
Entrypoints keep a separate client-side record for per-request output queues and metrics; that record is not a substitute for orchestrator state.

\begin{figure}[t]
  \centering
  \resizebox{0.42\linewidth}{!}{
\begin{tikzpicture}[
  x=1cm, y=1cm,
  font=\sffamily\scriptsize,
  flow arr/.style={
    omniInk, line width=0.7pt,
    -{Stealth[length=1.6mm, width=1.2mm]},
  },
  flow line/.style={omniInk, line width=0.7pt},
  flow box/.style={
    omni process,
    font=\sffamily\scriptsize,
    minimum width=4.4cm,
    minimum height=0.52cm,
    inner xsep=4pt,
    inner ysep=2pt,
  },
  flow box teal/.style={
    omni process teal,
    font=\sffamily\scriptsize,
    minimum width=2.45cm,
    minimum height=0.48cm,
    inner sep=2pt,
  },
]
  \node[flow box] (n1) at (0, 5.15)
    {\texttt{add\_request} / \texttt{streaming\_update}};
  \node[omni step, minimum size=3.8mm] at ([xshift=-2.6mm]n1.west) {1};

  \node[flow box] (n2) at (0, 4.25)
    {\texttt{create OrchestratorRequestState}};
  \node[omni step, minimum size=3.8mm] at ([xshift=-2.6mm]n2.west) {2};

  \node[flow box] (n3) at (0, 3.35)
    {\texttt{submit Stage 0}};
  \node[omni step, minimum size=3.8mm] at ([xshift=-2.6mm]n3.west) {3};

  \node[omni decision, aspect=2.6, inner sep=1.6pt,
        minimum width=2.8cm, font=\sffamily\footnotesize]
    (n4) at (0, 2.18) {\texttt{poll} stage outputs};
  \node[omni step teal, minimum size=3.8mm] at ([xshift=-2.8mm]n4.west) {4};

  \draw[flow arr] (n1) -- (n2);
  \draw[flow arr] (n2) -- (n3);
  \draw[flow arr] (n3) -- (n4);

  \node[flow box teal] (fout) at (-2.55, 0.72) {client output};
  \node[omni decision, aspect=2.35, inner sep=1.6pt,
        minimum width=2.45cm, font=\sffamily\footnotesize]
    (adv) at (2.55, 0.72) {more stages?};

  \coordinate (fork) at (0, 1.52);
  \draw[flow line] (n4.south) -- (fork);
  \draw[flow line] (fout.north |- fork) -- (adv.north |- fork);
  \draw[flow arr] (fork -| fout) -- (fout.north);
  \draw[flow arr] (fork -| adv) -- (adv.north);
  \node[font=\sffamily\tiny, text=omniGray, anchor=south]
    at ($($(fout.north |- fork)!0.5!(fork)$)+(0,0.06)$) {emit};
  \node[font=\sffamily\tiny, text=omniGray, anchor=south]
    at ($($(adv.north |- fork)!0.5!(fork)$)+(0,0.06)$) {check};

  \node[flow box teal] (async) at (2.55, -0.22) {\texttt{async\_chunk}};
  \node[flow box teal] (batch) at (2.55, -1.00) {\texttt{batch}};
  \draw[flow arr] (adv.south) -- (async.north);
  \draw[flow arr] (async.south) -- (batch.north);
  \node[font=\sffamily\tiny, text=omniGray, anchor=west]
    at ([xshift=2.2mm]$(adv.south)!0.50!(async.north)$) {yes};

  \draw[flow line] (adv.east) -- ++(0.38,0) coordinate (norail)
    -- (norail |- 0,-1.58);
  \node[font=\sffamily\tiny, text=omniGray, anchor=south]
    at ([yshift=1.4mm]$(adv.east)+(0.18,0)$) {no};

  \def\rail{-1.58}
  \draw[flow line] (fout.south) -- (fout.south |- 0,\rail);
  \draw[flow line] (batch.south) -- (batch.south |- 0,\rail);
  \draw[flow line] (fout.south |- 0,\rail) -- (norail |- 0,\rail);
  \node[font=\sffamily\tiny, text=omniGray, anchor=east]
    at ([xshift=-1.2mm]fout.south |- 0,\rail+0.14) {done};

  \node[flow box, minimum width=4.35cm] (done) at (0, -2.28)
    {final-output set done $\rightarrow$ cleanup};
  \draw[flow arr] (0, \rail) -- (done.north);

  \omniBalanceLR
\end{tikzpicture}}
  \caption{Orchestrator request flow: admit, poll, then branch to client delivery or stage advancement (\texttt{async\_chunk} or batch).}
  \label{fig:orchestrator-flow}
\end{figure}
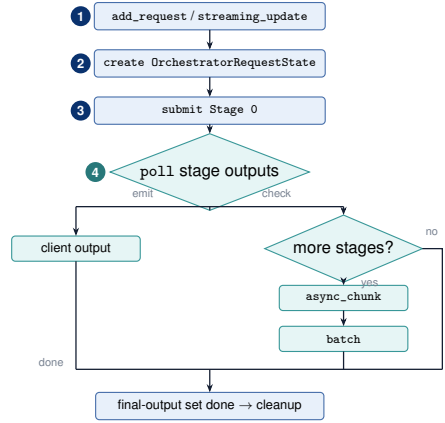

\paragraph{Advancement Loop.}
Upon admission, the orchestrator submits the request to stage~0 through the stage pool and then polls stage outputs.
An event-driven variant of this loop wakes on stage output instead of polling on a timer.
It is the default for the two-stage Qwen3-TTS pipeline and opt-in elsewhere.
Each polled result is routed along one or both of two paths
(Figure~\ref{fig:orchestrator-flow}).

The first path is client delivery.
Any stage marked as client-emitting may produce an incremental output for the entrypoint.
Multiple stages may emit concurrently, so partial results can reach users without requiring the entire pipeline to drain.
The second path is stage advancement.
In the batch path, the orchestrator constructs the next-stage request after an upstream stage finishes, or after a streaming segment finishes when the bridge needs earlier handoff, and submits it to a sticky downstream replica.
In the chunked path described below, downstream stages are pre-warmed and heavy payloads move on the connector data plane, so the orchestrator suppresses a redundant control-plane forward for those edges.
Replica selection itself is delegated to the \texttt{StagePool}: sticky affinity preserves streaming updates and in-stage KV locality, while unbound requests follow the pool's load-balancing policy
(Section~\ref{sec:arch-engines}).
Abort and cancellation reverse the same path.
An entrypoint abort becomes an orchestrator abort that releases replica bindings and clears request state, so stranded stage work does not continue after the client has left.

\paragraph{Inter-Stage Chunk Streaming.}
Many multi-stage pipelines do not need a complete upstream sequence before downstream computation can begin.
Speech and omni pipelines are representative: waveform or codec decoding can start as soon as early frames arrive.
\sys{} supports this overlap with \texttt{async\_chunk}.
When the mode is enabled, the orchestrator pre-submits placeholder requests on chunk-receiving stages so that workers can begin receiving immediately.
As the upstream stage produces tokens or codec frames, workers publish chunks through the connector and stage-specific chunk processors.
Downstream engines therefore decode in parallel with upstream generation, which is the primary mechanism for reducing time to first audio packet.
For those edges the orchestrator retains lifecycle ownership (admission, affinity, abort, and completion) but does not re-issue a full control-plane forward for every chunk.
The batch full-payload path remains available for pipelines that prefer coarser handoffs.
\sys{} does not force every edge to stream; the pipeline configuration selects the transfer mode per edge.

\paragraph{Client-Side Streaming as Request Progress.}
Client streaming is likewise modeled as request progress rather than as an out-of-band channel.
This is the one-shot streaming contract: input chunks and output deltas share a single request identity from admission to completion.
For streaming input, the entrypoint submits the first chunk as a resumable stage~0 request.
Subsequent chunks arrive as streaming updates on the same request identity.
An end-of-stream update clears the resumable flag and terminates further appends.
Unknown-id updates are dropped, so a late chunk cannot resurrect an aborted or unadmitted request.
The orchestrator tracks segment boundaries and the raw tokens required by stage bridges, because processed stage outputs alone may discard information needed for the next handoff.
For streaming output, any client-emitting stage can yield incremental outputs.
The entrypoint demultiplexes them onto per-request queues and yields them to adapters, so TTS, speech, and video serving share the same delivery path.
Stage transfer never returns through the client API; adapters observe only client-emitting stages.

Some diffusion workloads further accept a prompt update while generation is already running.
The entrypoint exposes this as an interaction submission on an active request, and the orchestrator forwards the event to stage~0 through the stage pool.
That path still lives inside one request's progress model.
When interaction must outlive a single generate---full-duplex speech, world-model rollouts, or robot loops---\sys{} adds session identity, retention, and admission on top of these same streaming hooks
(Section~\ref{sec:arch-session}).

\subsection{Execution Engine}
\label{sec:arch-engines}

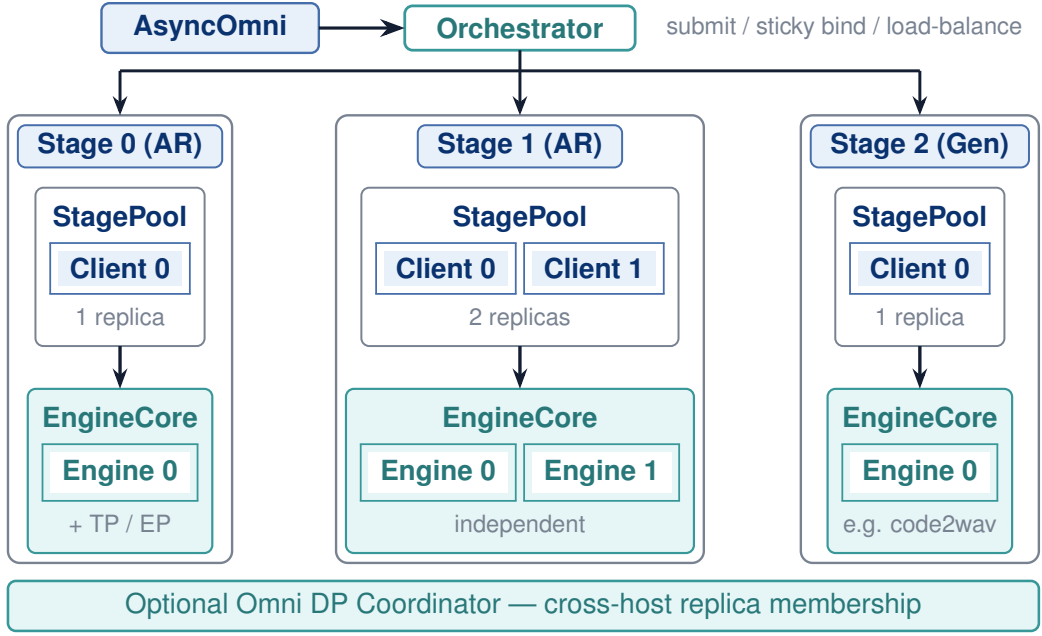
\begin{figure}[t]
  \centering
  \resizebox{\linewidth}{!}{
\begin{tikzpicture}[
  x=1cm, y=1cm,
  font=\sffamily,
  every node/.style={outer sep=0pt},
  stage title/.style={
    draw=omniBlueSoft, fill=omniBlueFill, text=omniBlue,
    rounded corners=2pt, line width=0.5pt,
    font=\sffamily\bfseries\scriptsize,
    inner xsep=5pt, inner ysep=2pt,
  },
  pool box/.style={
    draw=omniGray, fill=white, rounded corners=2pt, line width=0.5pt,
    align=center, inner xsep=3.5pt, inner ysep=2pt,
  },
  core box/.style={
    draw=omniTeal, fill=omniTealFill, rounded corners=2pt, line width=0.55pt,
    align=center, inner xsep=3.5pt, inner ysep=2pt,
  },
  colframe/.style={
    draw=omniGray, rounded corners=3pt, line width=0.6pt,
    inner xsep=2.5pt, inner ysep=2.5pt,
  },
]
  \newcommand{\clientchip}[1]{%
    \begingroup\setlength{\fboxsep}{1.8pt}\setlength{\fboxrule}{0.45pt}%
    {\color{omniBlueSoft}\fbox{\colorbox{omniBlueFill}{%
      \textcolor{omniBlue}{\sffamily\bfseries\scriptsize\,#1\,}}}}%
    \endgroup}
  \newcommand{\enginechip}[1]{%
    \begingroup\setlength{\fboxsep}{1.8pt}\setlength{\fboxrule}{0.45pt}%
    {\color{omniTeal}\fbox{\colorbox{white}{%
      \textcolor{omniTealDark}{\sffamily\bfseries\scriptsize\,#1\,}}}}%
    \endgroup}

  \def\xA{2.45}
  \def\xB{6.0}
  \def\xC{9.55}
  \def\yTitle{-1.05}
  \def\yRail{-0.38}

  \node[
    draw=omniBlueSoft, fill=omniBlueFill, text=omniBlue,
    rounded corners=2pt, line width=0.6pt,
    font=\sffamily\bfseries\scriptsize,
    inner xsep=8pt, inner ysep=3pt,
  ] (async) at (3.25, 0) {AsyncOmni};
  \node[
    draw=omniTeal, fill=white, text=omniTealDark,
    rounded corners=2pt, line width=0.7pt,
    font=\sffamily\bfseries\scriptsize,
    inner xsep=8pt, inner ysep=3pt,
  ] (orch) at (6.0, 0) {Orchestrator};
  \draw[omni ctrl] (async) -- (orch);
  \node[font=\sffamily\tiny, text=omniGray, anchor=west]
    at ($(orch.east)+(0.16,0)$) {submit / sticky bind / load-balance};

  \node[stage title] (s0t) at (\xA, \yTitle) {Stage 0 (AR)};
  \node[pool box, below=0.18cm of s0t] (poolA) {%
    \begin{tabular}{@{}c@{}}
      {\sffamily\bfseries\scriptsize\textcolor{omniBlue}{StagePool}}\\[0.8pt]
      \clientchip{Client 0}\\[0.4pt]
      {\sffamily\tiny\textcolor{omniGray}{1 replica}}
    \end{tabular}%
  };
  \node[core box, below=0.38cm of poolA] (coreA) {%
    \begin{tabular}{@{}c@{}}
      {\sffamily\bfseries\scriptsize\textcolor{omniTealDark}{EngineCore}}\\[0.8pt]
      \enginechip{Engine 0}\\[0.4pt]
      {\sffamily\tiny\textcolor{omniGray}{+ TP / EP}}
    \end{tabular}%
  };
  \draw[omniInk, line width=0.85pt, -{Stealth[length=1.8mm, width=1.3mm]}]
    (poolA.south) -- (coreA.north);
  \begin{scope}[on background layer]
    \node[colframe, fit=(s0t)(poolA)(coreA)] (colA) {};
  \end{scope}

  \node[stage title] (s1t) at (\xB, \yTitle) {Stage 1 (AR)};
  \node[pool box, below=0.18cm of s1t] (poolB) {%
    \begin{tabular}{@{}c@{}}
      {\sffamily\bfseries\scriptsize\textcolor{omniBlue}{StagePool}}\\[0.8pt]
      \clientchip{Client 0}\,\clientchip{Client 1}\\[0.4pt]
      {\sffamily\tiny\textcolor{omniGray}{2 replicas}}
    \end{tabular}%
  };
  \node[core box, below=0.38cm of poolB] (coreB) {%
    \begin{tabular}{@{}c@{}}
      {\sffamily\bfseries\scriptsize\textcolor{omniTealDark}{EngineCore}}\\[0.8pt]
      \enginechip{Engine 0}\,\enginechip{Engine 1}\\[0.4pt]
      {\sffamily\tiny\textcolor{omniGray}{independent}}
    \end{tabular}%
  };
  \draw[omniInk, line width=0.85pt, -{Stealth[length=1.8mm, width=1.3mm]}]
    (poolB.south) -- (coreB.north);
  \begin{scope}[on background layer]
    \node[colframe, fit=(s1t)(poolB)(coreB)] (colB) {};
  \end{scope}

  \node[stage title] (s2t) at (\xC, \yTitle) {Stage 2 (Gen)};
  \node[pool box, below=0.18cm of s2t] (poolC) {%
    \begin{tabular}{@{}c@{}}
      {\sffamily\bfseries\scriptsize\textcolor{omniBlue}{StagePool}}\\[0.8pt]
      \clientchip{Client 0}\\[0.4pt]
      {\sffamily\tiny\textcolor{omniGray}{1 replica}}
    \end{tabular}%
  };
  \node[core box, below=0.38cm of poolC] (coreC) {%
    \begin{tabular}{@{}c@{}}
      {\sffamily\bfseries\scriptsize\textcolor{omniTealDark}{EngineCore}}\\[0.8pt]
      \enginechip{Engine 0}\\[0.4pt]
      {\sffamily\tiny\textcolor{omniGray}{e.g. code2wav}}
    \end{tabular}%
  };
  \draw[omniInk, line width=0.85pt, -{Stealth[length=1.8mm, width=1.3mm]}]
    (poolC.south) -- (coreC.north);
  \begin{scope}[on background layer]
    \node[colframe, fit=(s2t)(poolC)(coreC)] (colC) {};
  \end{scope}

  \coordinate (rail) at (0, \yRail);
  \draw[omniInk, line width=0.75pt] (orch.south) -- (6.0, \yRail);
  \draw[omniInk, line width=0.75pt] (colA.north |- rail) -- (colC.north |- rail);
  \foreach \n in {colA, colB, colC} {
    \draw[omniInk, line width=0.75pt, -{Stealth[length=1.7mm, width=1.3mm]}]
      (\n.north |- rail) -- (\n.north);
  }

  \path
    let \p1=(colA.west), \p2=(colC.east),
        \p3=(colA.south), \p4=(colB.south), \p5=(colC.south) in
    node[
      draw=omniTeal, fill=omniTealFill, text=omniTealDark,
      rounded corners=2pt, line width=0.65pt,
      font=\sffamily\scriptsize,
      inner xsep=8pt, inner ysep=3pt,
      anchor=north,
      minimum width={veclen(\x2-\x1,0)},
    ] (coord) at ({0.5*(\x1+\x2)},{min(\y3,\y4,\y5)-0.16cm})
      {Optional Omni DP Coordinator --- cross-host replica membership};

  \omniBalanceLR
\end{tikzpicture}}
  \caption{Per-stage columns stack a StagePool (clients) over EngineCore replicas; Stage~1 shows scale-out. Sticky affinity is the vertical client--engine pairing. Optional Omni DP Coordinator covers cross-host membership.}
  \label{fig:engines-replicas}
\end{figure}

A multi-stage pipeline becomes executable only once each stage is bound to a compute engine and, when necessary, replicated for capacity.
Omni pipelines mix autoregressive decoding, non-autoregressive tensor generation, and iterative diffusion.
These paradigms impose incompatible scheduler assumptions if forced into one worker.
\sys{} therefore specializes engines by stage type and scales each logical stage through an independent replica pool
(Figure~\ref{fig:engines-replicas}).

\paragraph{Autoregressive and Generation Engines.}
Non-diffusion stages reuse the worker and scheduler stack of vLLM~\cite{vllm}, extended with Omni-specific runners and output processors.
Two worker roles cover the dominant non-diffusion patterns.
An autoregressive worker executes thinker- and talker-style stages with continuous batching, token sampling, and KV-cache management, and can export hidden states or pooled tensors as cross-stage artifacts.
A generation worker executes non-autoregressive codec or waveform stages such as code-to-wav: it returns tensor outputs for downstream stages or clients rather than vocabulary logits.
Both roles can refresh stage inputs from upstream state on each iteration, which is common when a talker repeatedly concatenates thinker context.
Each stage selects its model runner independently, using either the vLLM V1 runner or the experimental Model Runner V2 (MRv2), so a single pipeline can mix V1 and MRv2 stages.
MRv2 stages keep per-request sampling and generation state on the device, replay FULL CUDA graphs that can also cover declared tensor-only auxiliary outputs, and expose opt-in model hooks for custom sampling and for passing sampled-token embeddings to the next stage.
On CUDA, the default Qwen3-TTS deployment selects MRv2, while Qwen3-Omni, MiniCPM-o~4.5 turn mode, Higgs Audio v3, MOSS-TTS-Local, and CosyVoice3 use V1 by default and offer opt-in MRv2 profiles, and non-CUDA platforms remain on V1.
Concrete workers are selected by the active hardware platform, so CUDA, ROCm, Ascend NPU, and related backends share the same stage contract while swapping kernels and process launchers
(Section~\ref{sec:platforms-hw}).

\paragraph{Diffusion Engine.}
Diffusion stages run a dedicated engine organized around request admission, model execution, and optional step-level control.
A scheduler decides when a diffusion job may enter execution; the engine then drives DiT or flow-matching denoising for the bound pipeline.
Compatible jobs may share a batch, and selected pipelines further support continuous batching across denoising steps rather than one-job-at-a-time execution.
When streaming frames or mid-request cancellation are required, individual denoising steps become schedulable units.
Where the model needs durable attention state, the diffusion scheduler can own a paged KV pool on the worker instead of leaving cache layout entirely to pipeline-local code
(Section~\ref{sec:arch-kv}).
Accelerations such as TeaCache and Cache-DiT, sequence or CFG parallelism, and transformer offload or quantization attach inside this engine rather than in the orchestrator
(Section~\ref{sec:platforms-efficiency}).
Diffusion-specific optimizations therefore do not leak into AR stage scheduling, while diffusion stages still participate in the same orchestrator control plane.

\paragraph{AR--Diffusion Backend.}
A subset of world-model pipelines needs diffusion-style execution together with cross-request paged attention state.
For these workloads, \sys{} provides an AR--diffusion backend that extends the diffusion engine with a runner-local paged KV pool built on vLLM's block manager.
The backend retains session-scoped KV across chunked rollouts inside one model instance, unlike a default diffusion stage that treats each request as an independent denoising job.
This backend is optional and currently oriented toward AR--DiT world-model serving, including realtime tick-style sessions that keep the stage warm across interaction steps.
It does not replace cross-stage AR-to-DiT KV transfer between separately deployed AR and diffusion stages
(Section~\ref{sec:arch-kv}).

\paragraph{Stage Replication and Parallel Composition.}
Capacity scaling inside a stage is handled by a \texttt{StagePool}.
Each logical stage owns exactly one pool; the pool may contain multiple replica engines.
The pool binds a request to a replica with sticky affinity so that streaming updates, multi-turn appends, and in-stage KV locality remain on the same replica.
Across unbound requests, the pool applies a load-balancing policy such as round-robin or least queue length.
Replicas are independent engine instances and do not share a collective world across the pool.
Tensor parallel, data parallel, pipeline parallel, and expert parallel degrees are applied inside each replica.
A composable strategy overlay can declare these intra-replica axes together with a stage-replica count, after which the pool owns replica membership and routing.
Remote workers attach through the distributed coordinator when stages are placed across hosts.
Process launchers such as Ray remain an orthogonal mechanism inside a replica and are not a substitute for stage-level replication.

\subsection{Intermediate Data Management}
\label{sec:arch-kv}

\begin{figure}[t]
  \centering
  \resizebox{\linewidth}{!}{
\begin{tikzpicture}[
  x=1cm, y=1cm,
  font=\sffamily,
]
  \node[omni teal layer, minimum width=2.9cm, minimum height=2.85cm]
    (c1) at (1.7, 2.55) {};
  \node[font=\sffamily\bfseries\footnotesize, text=omniTealDark]
    at (1.7, 3.65) {Hidden states};
  \node[font=\sffamily\scriptsize, text=omniInk, align=center, text width=2.5cm]
    at (1.7, 3.15) {activations /\\embeddings};
  \node[omni teal card, minimum width=2.4cm, minimum height=0.7cm]
    at (1.7, 2.4) {\scriptsize Thinker$\rightarrow$Talker};
  \node[font=\sffamily\tiny, text=omniGray, align=center]
    at (1.7, 1.7) {\texttt{async\_chunk}};

  \node[omni teal layer, minimum width=2.9cm, minimum height=2.85cm]
    (c2) at (4.85, 2.55) {};
  \node[font=\sffamily\bfseries\footnotesize, text=omniTealDark]
    at (4.85, 3.65) {Codec codes};
  \node[font=\sffamily\scriptsize, text=omniInk, align=center, text width=2.5cm]
    at (4.85, 3.15) {RVQ / discrete\\audio tokens};
  \node[omni teal card, minimum width=2.4cm, minimum height=0.7cm]
    at (4.85, 2.4) {\scriptsize Talker$\rightarrow$Code2Wav};
  \node[font=\sffamily\tiny, text=omniGray, align=center]
    at (4.85, 1.7) {chunk + context};

  \node[omni layer, minimum width=2.9cm, minimum height=2.85cm]
    (c3) at (8.0, 2.55) {};
  \node[font=\sffamily\bfseries\footnotesize, text=omniBlue]
    at (8.0, 3.65) {KV blocks};
  \node[font=\sffamily\scriptsize, text=omniInk, align=center, text width=2.5cm]
    at (8.0, 3.15) {paged K/V\\block table};
  \node[omni card, minimum width=2.4cm, minimum height=0.7cm]
    at (8.0, 2.4) {\scriptsize PD / AR$\rightarrow$DiT};
  \node[font=\sffamily\tiny, text=omniGray, align=center]
    at (8.0, 1.7) {KV transfer};

  \node[omni layer, minimum width=2.9cm, minimum height=2.85cm]
    (c4) at (11.15, 2.55) {};
  \node[font=\sffamily\bfseries\footnotesize, text=omniBlue]
    at (11.15, 3.65) {Latents};
  \node[font=\sffamily\scriptsize, text=omniInk, align=center, text width=2.5cm]
    at (11.15, 3.15) {DiT / VAE\\tensors};
  \node[omni card, minimum width=2.4cm, minimum height=0.7cm]
    at (11.15, 2.4) {\scriptsize DiT$\rightarrow$VAE};
  \node[font=\sffamily\tiny, text=omniGray, align=center]
    at (11.15, 1.7) {stage output};

  \node[omni connector box, minimum width=12.6cm, minimum height=0.55cm]
    (conn) at (6.4, 0.7) {OmniConnector};
  \foreach \n in {c1,c2,c3,c4} {
    \draw[omniInk, line width=0.65pt, dashed, dash pattern=on 2pt off 1.5pt]
      (\n.south) -- (\n.south |- conn.north);
  }

  \omniBalanceLR
\end{tikzpicture}}
  \caption{Four inter-stage payload kinds over OmniConnector.}
  \label{fig:kv-paths}
\end{figure}
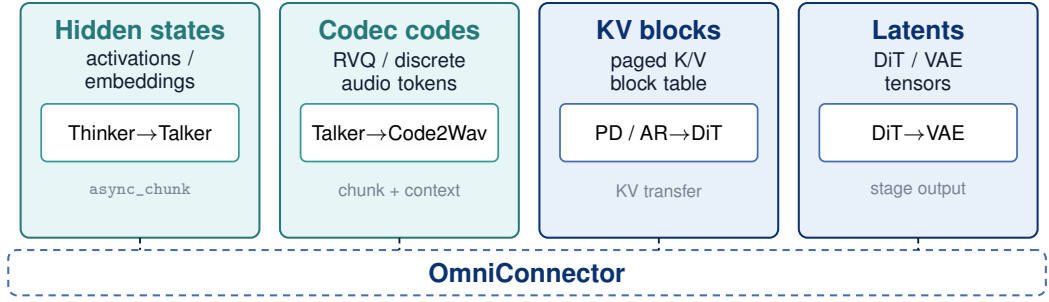

Disaggregated stages must exchange typed intermediate state without shipping every tensor through the orchestrator process.
That state is broader than attention KV.
Qwen3-Omni is representative: the thinker exports hidden states and embeddings to the talker; the talker exports discrete codec codes to the vocoder.
Diffusion pipelines add latents between DiT and VAE stages.
Prefill--decode and AR-to-DiT edges additionally move paged KV blocks extracted from a stage's block table.
\sys{} therefore treats intermediate-data management as the discipline of \emph{what} each edge carries and \emph{when} it becomes available, while OmniConnector (Section~\ref{sec:arch-connector}) supplies the transport that moves the bytes
(Figure~\ref{fig:kv-paths}).

\paragraph{Payload Kinds.}
Inter-stage outputs are typed payloads assembled by stage runners and bridge processors, not opaque blobs.
The dominant kinds are:
\begin{enumerate}[leftmargin=1.6em,itemsep=0.35em,parsep=0pt]
  \item Hidden states and embeddings condition a downstream AR stage. On Qwen3-Omni, the thinker$\rightarrow$talker edge carries selected-layer hidden states, prompt embeddings, and TTS special embeddings, together with token-id metadata used to size the talker prompt.
  \item Codec codes (for example RVQ rows) feed a generation-stage vocoder. The talker$\rightarrow$code2wav edge publishes \texttt{codes.audio} with chunk-boundary metadata such as left-context size and finished flags.
  \item Paged KV blocks move on prefill--decode and AR$\rightarrow$DiT edges as key/value tensors extracted from vLLM's paged block table. These are attention-cache blocks, not hidden-state activations.
  \item Latents and multimodal tensors move on diffusion and world-model edges (DiT/VAE latents and related side tensors that are neither AR activations nor KV blocks).
\end{enumerate}
Client-visible outputs are partitioned from inter-stage payloads so that adapters see only stages marked as client-emitting.
Bridge processors decide how each edge packs its payload (chunked versus full) and what placeholder the orchestrator may submit on the control plane.

\paragraph{Transfer Modes.}
Two delivery modes cover most pipelines.
Under \texttt{async\_chunk}, producers publish payload chunks as soon as tokens or frames are ready; consumers are pre-warmed and receive on the connector data plane while upstream generation continues.
Under full-payload transfer, the producer accumulates until the stage finishes, then publishes once; the consumer may wait behind a scheduling gate until the payload arrives.
Both modes use the same connector put/get surface.
They differ in lifecycle and in how much control-plane forwarding the orchestrator suppresses for that edge.
Speech defaults lean on chunking so waveform decoding overlaps codec prediction; coarser pipelines keep the batch path.
On some TTS paths, an adaptive chunk ramp further adjusts chunk size from downstream buffer feedback rather than fixing one watermark for the whole utterance.

\paragraph{KV as a Specialized Payload.}
KV movement reuses the connector fabric but follows a dedicated manager rather than the hidden-state/codec chunk path.
For prefill--decode disaggregation, adjacent AR stages are marked producer and consumer under a KV-transfer connector such as Mooncake~\cite{mooncake,epd}.
The orchestrator injects remote bootstrap and request identifiers into decode sampling parameters, and retains multimodal side outputs that must survive the handoff so omni PD is not text-only.
For AR$\rightarrow$DiT, an Omni KV transfer manager extracts finished AR blocks, publishes them under KV-specific keys, and the diffusion runner injects them as \texttt{past\_key\_values} before denoising.
In both cases the bytes still move on OmniConnector (or the configured KV backend); only the extraction, keying, and apply steps are KV-specific.
Intra-stage decode continues to use the stage-local paged KV manager for ordinary attention, including optional prefix reuse of cached hidden and multimodal side tensors on a block hit.
Diffusion workers can likewise allocate and hold paged KV under scheduler-managed block tables for models that need durable attention state on the DiT side (for example HunyuanImage3), so diffusion cache ownership is no longer only ad-hoc model-local buffers.
That destination-side pool is what lets AR$\rightarrow$DiT transfer land into reserved blocks instead of opaque contiguous dumps; coverage still varies by pipeline.

\paragraph{Boundary with the Connector.}
Intermediate-data management answers \emph{which} tensors an edge needs and \emph{how} producers and consumers agree on readiness.
The connector answers \emph{where} those bytes live and \emph{how} they are copied between processes or hosts.
The orchestrator retains admission, sticky routing, abort, and completion; it never becomes a bulk tensor bus.
Section~\ref{sec:arch-connector} develops the transport contract and backend selection that make this split concrete.

\subsection{Connector}
\label{sec:arch-connector}

Section~\ref{sec:arch-kv} classifies the intermediate payloads stages exchange.
This subsection covers the transport that moves those payloads once an edge has decided what to send.
Hidden states, codec chunks, latents, and packed KV blocks can be large enough that shipping them through the orchestrator process would serialize the control loop and inflate host memory traffic.
At the same time, the orchestrator must retain a precise view of request membership, cancellation, and completion.
\sys{} resolves this tension with an explicit control/data split.
Scheduling signals and lightweight metadata remain on the orchestrator path.
Heavy payloads move through OmniConnector on the data plane shown in the
system overview (Section~\ref{sec:arch-overview}).

\paragraph{Transport Contract.}
OmniConnector exposes a narrow transport interface for put, get, cleanup, and health checks.
A producer publishes a payload under a key and returns connector metadata small enough to travel on the control queue, such as a shared-memory name or a remote transfer handle.
The consumer uses that metadata to fetch the payload into local buffers.
The connector itself is semantics-agnostic: it does not interpret whether a blob is a hidden-state chunk, a codec frame, or a KV block.
Chunk-transfer and KV-transfer layers sit above this interface and own request-level lifecycle: when keys are created, retried, prefetched, or retired.
The same transport can therefore carry stage inputs, \texttt{async\_chunk} frames, and KV blobs without proliferating per-artifact networking stacks.

\paragraph{Placement-Aware Backends.}
Backend selection follows placement.
On a single node, shared memory is the default path for large tensors because it avoids copies through the orchestrator address space while preserving low-latency access between local stage processes.
Across nodes, \sys{} supports store-oriented and transfer-engine backends, including Mooncake TCP and RDMA paths, together with platform-specific transports where required by the accelerator fabric~\cite{mooncake}.
Edges between stages may choose backends independently.
A local thinker-to-talker edge can remain on shared memory for hidden-state chunks, while a prefill-to-decode or AR-to-remote-DiT edge uses a transfer-engine backend for KV blocks.
Health and cleanup hooks allow failed transfers to be detected without leaving orphaned segments in shared memory or remote stores.

\paragraph{Composition with Streaming and KV Movement.}
Connector usage is shared by chunk streaming and KV transfer, but their control policies differ.
Under \texttt{async\_chunk}, workers publish chunk keys as upstream tokens or frames become available, and the orchestrator skips redundant control-plane forwards for edges that already receive chunks on the data plane
(Section~\ref{sec:arch-entrypoint-orch}).
Under AR-to-DiT or PD KV transfer, the KV path packs blocks and often prefers transfer-engine backends for bulk movement, optionally prefetching before the consumer blocks on context injection
(Section~\ref{sec:arch-kv}).
In both settings the invariant is identical.
Admission, routing, and completion stay with the orchestrator.
Payload bytes stay with the connector.
That invariant is what keeps multi-stage disaggregation practical as pipelines grow beyond a single host.

\subsection{Session-Oriented Control}
\label{sec:arch-session}

Streaming in Section~\ref{sec:arch-entrypoint-orch} describes progress \emph{inside} one request: chunked inputs, incremental outputs, and inter-stage \texttt{async\_chunk} handoffs share a single request identity until the final-output set drains.
A growing fraction of omni workloads need something stronger.
Full-duplex speech agents continue listening while synthesizing responses.
World models retain latent state across rollouts.
Streaming video sessions accept mid-generation edits.
Robot policies repeatedly map observations to actions under a stable interaction identity.
These workloads still execute on the multi-stage runtime, but they cannot be described as one-shot generate-and-complete jobs.

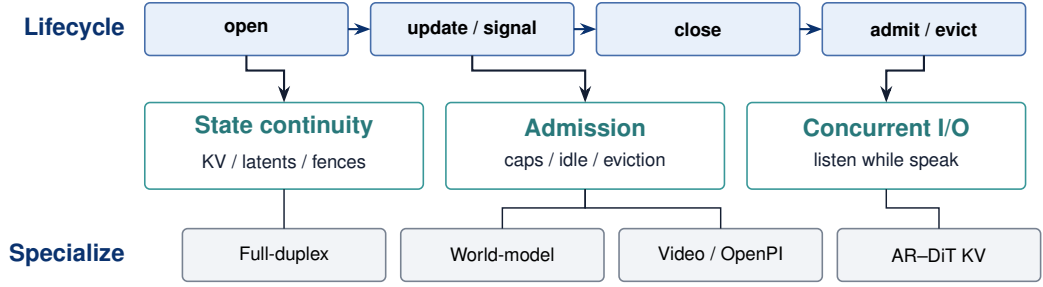
\begin{figure}[t]
  \centering
  \resizebox{\linewidth}{!}{
\begin{tikzpicture}[
  x=1cm, y=1cm,
  font=\sffamily,
]
  \node[omni card, fill=omniBlueFill, minimum width=2.7cm, minimum height=0.7cm]
    (l1) at (1.7, 3.55) {\scriptsize\textbf{open}};
  \node[omni card, fill=omniBlueFill, minimum width=2.7cm, minimum height=0.7cm]
    (l2) at (4.7, 3.55) {\scriptsize\textbf{update / signal}};
  \node[omni card, fill=omniBlueFill, minimum width=2.7cm, minimum height=0.7cm]
    (l3) at (7.7, 3.55) {\scriptsize\textbf{close}};
  \node[omni card, fill=omniBlueFill, minimum width=2.7cm, minimum height=0.7cm]
    (l4) at (10.7, 3.55) {\scriptsize\textbf{admit / evict}};
  \draw[omni arrow] (l1) -- (l2);
  \draw[omni arrow] (l2) -- (l3);
  \draw[omni arrow] (l3) -- (l4);
  \node[font=\sffamily\bfseries\footnotesize, text=omniBlue, anchor=east]
    at (0.2, 3.55) {Lifecycle};

  \node[omni teal card, minimum width=3.7cm, minimum height=1.15cm]
    (st) at (2.2, 2.0) {};
  \node[font=\sffamily\bfseries\footnotesize, text=omniTealDark]
    at (2.2, 2.25) {State continuity};
  \node[font=\sffamily\scriptsize, text=omniInk]
    at (2.2, 1.8) {KV / latents / fences};

  \node[omni teal card, minimum width=3.7cm, minimum height=1.15cm]
    (ad) at (6.2, 2.0) {};
  \node[font=\sffamily\bfseries\footnotesize, text=omniTealDark]
    at (6.2, 2.25) {Admission};
  \node[font=\sffamily\scriptsize, text=omniInk]
    at (6.2, 1.8) {caps / idle / eviction};

  \node[omni teal card, minimum width=3.7cm, minimum height=1.15cm]
    (io) at (10.2, 2.0) {};
  \node[font=\sffamily\bfseries\footnotesize, text=omniTealDark]
    at (10.2, 2.25) {Concurrent I/O};
  \node[font=\sffamily\scriptsize, text=omniInk]
    at (10.2, 1.8) {listen while speak};

  \draw[omni ctrl] (l1.south) -- ++(0,-0.2) -| (st.north);
  \draw[omni ctrl] (l2.south) -- ++(0,-0.2) -| (ad.north);
  \draw[omni ctrl] (l4.south) -- ++(0,-0.2) -| (io.north);

  \node[omni gray card, minimum width=2.7cm, minimum height=0.7cm]
    (s1) at (2.2, 0.55) {\scriptsize Full-duplex};
  \node[omni gray card, minimum width=2.7cm, minimum height=0.7cm]
    (s2) at (5.1, 0.55) {\scriptsize World-model};
  \node[omni gray card, minimum width=2.7cm, minimum height=0.7cm]
    (s3) at (8.0, 0.55) {\scriptsize Video / OpenPI};
  \node[omni gray card, minimum width=2.7cm, minimum height=0.7cm]
    (s4) at (10.9, 0.55) {\scriptsize AR--DiT KV};
  \node[font=\sffamily\bfseries\footnotesize, text=omniBlue, anchor=east]
    at (0.2, 0.55) {Specialize};
  \draw[omniInk, line width=0.55pt]
    (st.south) -- ++(0,-0.25) -| (s1.north);
  \draw[omniInk, line width=0.55pt]
    (ad.south) -- ++(0,-0.25) -| (s2.north);
  \draw[omniInk, line width=0.55pt]
    (ad.south) -- ++(0,-0.25) -| (s3.north);
  \draw[omniInk, line width=0.55pt]
    (io.south) -- ++(0,-0.25) -| (s4.north);

  \omniBalanceLR
\end{tikzpicture}}
  \caption{Session control: lifecycle steps, three shared concerns, and modality specializations.}
  \label{fig:session-control}
\end{figure}

\sys{} therefore treats session-oriented control as an extension of the same orchestrator, not as a second runtime
(Figure~\ref{fig:session-control}).
Streaming supplies the I/O hooks; the session layer adds a longer-lived identity, retention policy, and capacity admission across turns or forward calls.
The concrete APIs differ by modality, yet the underlying control requirements recur.

\paragraph{Session Lifecycle.}
A session carries an identity that outlives a single generate invocation.
It can be opened, updated, signaled, and closed.
Capacity admission may reject new sessions when a deployment profile is saturated, which is essential for realtime speech paths whose per-session GPU footprint is large.
Idle leases, explicit reset, and eviction reclaim resources when clients disconnect or stop producing useful work.
Full-duplex control uses open, append, signal, and close operations with hard session caps and fencing metadata.
World-model state managers create typed sessions and reclaim them under LRU policy when capacity is exceeded.
Video and robot entrypoints expose start, interaction or infer, and stop or reset operations with idle timeouts.
Across these surfaces, lifecycle policy belongs to the control plane; stage engines remain responsible for compute.

\paragraph{State Continuity.}
Sessions exist because some state must persist across turns and forward calls.
That state may include conversation or stage KV, latent buffers, epoch and turn fences, or engine-owned AR--diffusion session KV.
Ownership is deliberate.
Paged KV used by AR--diffusion runners remains inside the engine KV pool, where block allocation and attention kernels already operate.
Non-KV world-model artifacts such as latent rings are managed by a typed session-state layer that accounts for device-resident bytes and supports allocate, commit, view, reset, and evict operations.
Fencing metadata prevents a delayed update from an older turn from mutating a newer response.
This split keeps paged KV and application-level session memory composable instead of collapsing every persistent tensor into one opaque bag.

\paragraph{Concurrent Bidirectional I/O.}
Session workloads commonly accept input while output is still in flight.
Duplex speech continues to append microphone audio during playback.
Streaming video can ingest frames or prompt interactions while denoising proceeds.
Robot serving repeatedly consumes observations and emits action chunks under a sticky session identifier.
These patterns reuse the streaming-update and interaction hooks from Section~\ref{sec:arch-entrypoint-orch}, then add session identity, retention, and admission on top.
Long-lived interaction therefore does not require a second orchestrator; it strengthens lifetime and concurrency semantics on the existing control plane.

\paragraph{Workload Specialization.}
Full-duplex control and world-model session state are evolving subsystems and should be read as capability layers rather than as universally enabled defaults.
AR--diffusion session KV is likewise specialized and is best understood together with the engine and intermediate-data discussions in Sections~\ref{sec:arch-engines} and~\ref{sec:arch-kv}.
Streaming video and OpenPI robot loops are thinner serving-facing specializations of the same axes.
Wire protocols for all of these surfaces are specified in Section~\ref{sec:models-apis}.
The architectural claim is correspondingly narrow: \sys{} lifts long-lived multimodal interaction into a shared session control model, instead of maintaining a separate runtime per interaction style.

\section{Platforms and Ecosystem}
\label{sec:platforms}

The multi-stage architecture in Section~\ref{sec:architecture} is useful in production only when it travels with the hardware, efficiency tooling, and surrounding workflows that operators already depend on.
This section develops three supporting layers: platform plugins for multiple accelerators, an efficiency stack that attaches to stages without changing the control plane, and ecosystem hooks for deployment recipes, upstream vLLM alignment, and RL training handoff.
Quantitative speedups are deferred to Section~\ref{sec:eval}.

\subsection{Hardware Platforms}
\label{sec:platforms-hw}

Hard-coding a device software stack into the orchestrator couples model support to a single accelerator line.
In practice the same Omni or diffusion pipeline must land on Hopper GPUs in one region and Ascend NPUs in another, without a second control plane.
\sys{} therefore follows the platform-plugin discipline of upstream vLLM~\cite{vllm}: exactly one \texttt{OmniPlatform} is resolved at process start from builtin detectors and optional out-of-tree entry points.
Detection probes the local device stack (for example NVML, AMDSMI, \texttt{torch.npu}, or XPU availability) and returns a platform class path.
Out-of-tree plugins can register additional platforms through the Omni entry-point group without forking the orchestrator.
After resolution, the orchestrator, stage, and connector contracts remain unchanged.
Only worker bindings, attention backends, and device preparation hooks swap.

\paragraph{Platform Contract.}
An \texttt{OmniPlatform} inherits vLLM's device platform and adds Omni-specific hooks that stages consult at construction time.
The required surface is small.
Each platform names its autoregressive and generation worker classes, and a default root for stage configuration packages.
Optional hooks cover the places where accelerators diverge most in Omni workloads: diffusion attention backend selection, fused-MoE runtime preparation, packed-module mappings for quantized checkpoints, diffusion worker configuration, and one-shot model patches that cannot live in a portable graph.
Because these hooks sit behind a single interface, a stage YAML can stay device-agnostic while the platform supplies the concrete kernels and runners.

\paragraph{Specialization Spectrum.}
Builtin backends differ in how deeply they specialize, not in whether they speak a different orchestration language.
CUDA is the default path: it reuses the shared GPU AR and generation workers and carries the broadest diffusion attention and quantization coverage, including compute-capability-aware routing among FlashAttention, cuDNN, and related backends.
ROCm and MUSA also reuse the shared GPU workers, but inject platform patches and attention defaults (for example Triton attention on ROCm, and a \texttt{vllm-musa} plugin dependency for MUSA).
Ascend NPU and Intel XPU take the opposite end of the spectrum: dedicated AR and generation workers, vendor collective backends (\texttt{hccl}, \texttt{xccl}/\texttt{ccl}), and platform-local runtime hooks.
NPU further illustrates why a plugin boundary matters for Omni.
TTS code2wav and tokenizer paths, Ascend custom ops, NZ weight formats, fused-MoE forward context, and even NPU-local connector variants are applied inside the platform rather than threaded through the orchestrator.
XPU follows a similar dedicated-worker pattern and is commonly deployed through the documented container flow.
Per-backend coverage continues to evolve with releases.

\paragraph{Portability Invariant.}
Porting a stage therefore reduces to selecting or implementing the platform hooks and pointing stage configuration at the resulting workers.
Omni and diffusion pipelines do not need a second orchestrator per device.
Heterogeneous clusters can still place different stages on different backends when topology or capacity requires it, because the orchestrator and connector already treat placement as a binding problem.
The invariant is that control-plane semantics stay device-agnostic while device-specific performance work remains inside platform plugins.

\subsection{Efficiency Stack}
\label{sec:platforms-efficiency}

Omni pipelines are memory- and latency-sensitive for different reasons than text-only LLM serving.
Diffusion transformers amplify weight footprint across many denoising steps; multi-stage omni models keep several engines resident; and speech paths care about first-packet latency as much as steady-state throughput.
\sys{} therefore exposes an efficiency stack that attaches to stages instead of replacing the runtime.
The same entrypoint and orchestrator continue to drive requests.
Stages opt into quantization, offload, compilation, and cache backends according to model and device constraints.

\paragraph{Quantization.}
Checkpoint and online quantization are routed through a unified factory that extends the vLLM quant registry with Omni-specific configurations.
Supported families include FP8 and INT8 weight-activation schemes, ModelOpt FP8 and NVFP4 checkpoints, GGUF, AutoRound and INC style checkpoint quantization, offline SVDQuant-style W4A4 packs for selected diffusion models, and CUDA-only bitsandbytes weight-only modes.
Ascend adds MXFP8, MXFP4, and msModelSlim paths for selected diffusion models.
Coverage is intentionally scoped.
Diffusion quantization focuses on the transformer body by default, leaving tokenizers, VAEs, and schedulers on the base precision unless a method guide says otherwise.
Multi-stage omni and TTS pipelines typically quantize the autoregressive thinker or language-model stage when the checkpoint carries a supported configuration; non-AR stages remain in higher precision until explicitly validated.
Unverified backend and model cells should be treated as experimental rather than universally available.

\paragraph{Offload and Sleep.}
Diffusion stages can move inactive modules or layers to CPU to reduce peak device memory during long denoise trajectories.
Host-weight plans and related layerwise offload paths further let large DiT deployments keep inactive weights off device without changing the orchestrator.
A separate sleep mode releases resident weights between serving and training phases, which is particularly relevant for RL colocation.
Level-1 sleep offloads weights to CPU and can be woken for subsequent serve intervals.
The same sleep/wake surface is available on AR stages through AsyncOmni for online serving suspension
(Section~\ref{sec:arch-entrypoint-orch}).
Deeper demapping hooks exist for more aggressive memory reclamation, but wake and reload semantics for the deepest level remain incomplete; operators should treat them as partial control-plane hooks rather than a finished memory manager.

\paragraph{Compilation and Diffusion Caches.}
Autoregressive stages inherit cudagraph and compilation behavior from upstream vLLM on platforms that support those paths.
Codec and vocoder stages can also capture model-owned CUDA graphs over bounded shape buckets and fall back to eager execution outside them.
Examples are the Qwen3-TTS Code2Wav and fused streaming decoders and, in opt-in profiles, CosyVoice3 HiFT decoding, the MOSS-TTS-Local codec, and Higgs Audio v3 codec shapes.
Diffusion stages default to regional \texttt{torch.compile} over hot blocks, which reduces kernel-launch overhead without requiring a single static full graph for every resolution and step count.
Full-graph compilation remains more brittle and is not the default.
For iterative denoisers, TeaCache and Cache-DiT provide first-class step-to-step reuse.
Secondary caches such as MagCache and model-specific step caches exist in tree, but they are less mature than the primary backends.
Parallelism knobs for diffusion (sequence parallel, CFG parallel, tensor and pipeline parallel, HSDP, expert parallel, and VAE parallel) compose with these caches under documented compatibility constraints, again without changing orchestrator control flow.

Collectively, these mechanisms let a deployment trade memory, latency, and throughput while preserving one entrypoint and one orchestrator.
They are accelerators on top of Section~\ref{sec:architecture}, not alternate runtimes.

\subsection{Ecosystem and RL}
\label{sec:platforms-ecosystem}

A serving runtime becomes an ecosystem component only when it is easy to deploy, easy to embed, and easy for adjacent systems to control.
\sys{} therefore invests in recipes and application clients for operators, release alignment with upstream vLLM for maintainability, and an explicit training--serving control surface for RL frameworks.

\paragraph{Recipes and Application Clients.}
Deployment recipes follow a vendor-organized layout aligned with the broader vLLM recipe ecosystem.
Each recipe documents a concrete model, a hardware sketch, and a serve command so that common Omni, TTS, image, video, world-model, and robot setups can be reproduced without reverse-engineering stage YAML from scratch.
On the application side, a ComfyUI integration provides thin nodes that call OpenAI-compatible image, video, understanding, and TTS endpoints.
The integration is intentionally a client of the Omni server rather than an in-process fork of the engine, which keeps workflow UI concerns outside the runtime boundary.

\paragraph{Upstream vLLM Alignment.}
\sys{} tracks the upstream vLLM major and minor release line.
Omni model registration plugs into vLLM's general plugin mechanism, and platform-specific dependency sets are selected by target device at build or install time.
Version skew between \sys{} and vLLM is detected at import time and reported explicitly.
This alignment allows Omni to reuse AR scheduling, paged KV management, and cudagraph infrastructure instead of maintaining a divergent LLM engine.
It also constrains Omni feature work to remain composable with upstream improvements rather than patching around a frozen fork.

\begin{figure}[t]
  \centering
  \resizebox{\linewidth}{!}{
\begin{tikzpicture}[
  x=1cm, y=1cm,
  font=\sffamily\small,
  >=Stealth,
]
  \def\W{12.4}
  \def\cx{6.2}
  \def\cardW{3.50}
  \def\engW{4.15}


  \node[
    draw=omniTeal, fill=omniTealFill, line width=0.7pt, rounded corners=3pt,
    minimum width=\W cm, minimum height=2.40cm,
  ] (trainers) at (\cx, 8.05) {};
  \node[font=\sffamily\bfseries\small, text=omniTealDark]
    at ($(trainers.north)+(0,-0.26)$)
    {Diffusion / Omni RL trainers (VeRL-Omni)};

  \node[omni card, draw=omniOrange, line width=0.65pt,
        minimum width=\cardW cm, minimum height=1.40cm]
    (flow) at (2.40, 7.85) {};
  \node[font=\sffamily\bfseries\small, text=omniOrange, align=center]
    at (2.40, 7.85)
    {FlowGRPO\\[1pt]{\normalfont\small (On-policy / async)}};

  \node[omni card, draw=omniBlueSoft, line width=0.65pt,
        minimum width=\cardW cm, minimum height=1.40cm]
    (ppo) at (\cx, 7.85) {};
  \node[font=\sffamily\bfseries\small, text=omniBlue, align=center]
    at (\cx, 7.85)
    {Omni-PPO\\[1pt]{\normalfont\small (On-policy / async)}};

  \node[omni card, draw=omniTeal, line width=0.65pt,
        minimum width=\cardW cm, minimum height=1.40cm]
    (dpo) at (10.00, 7.85) {};
  \node[font=\sffamily\bfseries\small, text=omniTealDark, align=center]
    at (10.00, 7.85)
    {DPO\\[1pt]{\normalfont\small (Online / offline)}};

  \node[
    draw=omniOrange, fill=omniOrangeFill, line width=0.7pt, rounded corners=3pt,
    minimum width=\W cm, minimum height=2.70cm,
  ] (sys) at (\cx, 4.55) {};
  \node[font=\sffamily\bfseries\small, text=omniOrange]
    at ($(sys.north)+(0,-0.26)$)
    {Training system (VeRL $\times$ VeRL-Omni)};

  \node[omni card, fill=white, draw=omniBlueSoft, line width=0.65pt,
        minimum width=\engW cm, minimum height=1.60cm]
    (actor) at (2.85, 4.45) {};
  \node[font=\sffamily\bfseries\small, text=omniBlue]
    at (2.85, 4.85) {Actor engine};
  \node[font=\sffamily\small, text=omniInk, align=center]
    at (2.85, 4.17) {DiffusersFSDP2\\[-0.5pt]VeOmni / Megatron};

  \node[omni card, fill=omniTealFill, draw=omniTeal, line width=0.65pt,
        minimum width=\engW cm, minimum height=1.60cm]
    (tq) at (9.55, 4.45) {};
  \node[font=\sffamily\bfseries, text=omniTealDark, align=center]
    at (9.55, 4.45) {TransferQueue\\[-1pt]/ RPC};

  \draw[omni arrow]
    (tq.west) --
    node[midway, above, font=\sffamily\footnotesize, text=omniInk, inner sep=1pt]
      {data samples}
    (actor.east);

  \draw[omni ctrl] (sys.north) -- (trainers.south);

  \node[
    draw=omniTeal, fill=omniTealFill, line width=0.7pt, rounded corners=3pt,
    minimum width=\W cm, minimum height=2.90cm,
  ] (gen) at (\cx, 0.85) {};
  \node[font=\sffamily\bfseries\small, text=omniTealDark]
    at ($(gen.north)+(0,-0.26)$)
    {Multi-modal generation and reward};

  \node[omni card, draw=omniOrange, fill=white, line width=0.7pt,
        minimum width=\engW cm, minimum height=1.85cm]
    (roll) at (2.85, 0.72) {};
  \node[font=\sffamily\bfseries\small, text=omniOrange]
    at (2.85, 1.28) {Rollout engine};
  \node[
    draw=omniTeal, fill=omniTealFill, line width=0.6pt, rounded corners=2pt,
    minimum width=3.55cm, minimum height=0.68cm,
    font=\sffamily\bfseries\small, text=omniTealDark,
  ] at (2.85, 0.55) {vLLM-Omni (vLLM)};

  \node[omni card, draw=omniBlueSoft, fill=white, line width=0.7pt,
        minimum width=\engW cm, minimum height=1.85cm]
    (rew) at (9.55, 0.72) {};
  \node[font=\sffamily\bfseries\small, text=omniInk]
    at (9.55, 1.28) {Reward engine};
  \node[
    draw=omniTeal, fill=omniTealFill, line width=0.55pt, rounded corners=6pt,
    minimum width=1.80cm, minimum height=0.55cm,
    font=\sffamily\small, text=omniTealDark,
  ] at (8.50, 0.55) {Visual};
  \node[
    draw=omniTeal, fill=omniTealFill, line width=0.55pt, rounded corners=6pt,
    minimum width=1.80cm, minimum height=0.55cm,
    font=\sffamily\small, text=omniTealDark,
  ] at (10.60, 0.55) {Audio};

  \draw[omni arrow]
    (roll.east) --
    node[midway, above, font=\sffamily\footnotesize, text=omniInk, align=center,
         inner sep=1pt]
      {generate\\[-1pt]trajectories}
    (rew.west);

  \draw[omni arrow] (actor.south) -- (roll.north);
  \node[font=\sffamily\footnotesize, text=omniInk, anchor=east, align=right]
    at ([xshift=-3.5mm]$(sys.south -| actor)!0.50!(gen.north -| roll)$)
    {update\\[-1pt]weights};

  \draw[omni arrow] (rew.north) -- (tq.south);
  \node[font=\sffamily\footnotesize, text=omniInk, anchor=west, align=left]
    at ([xshift=3.5mm]$(gen.north -| rew)!0.50!(sys.south -| tq)$)
    {trajectories\\[-1pt]\& feedback};

  \omniBalanceLR
\end{tikzpicture}}
  \caption{\sys{} as the multi-modal rollout backend in VeRL-Omni.}
  \label{fig:verl_omni_arch}
\end{figure}
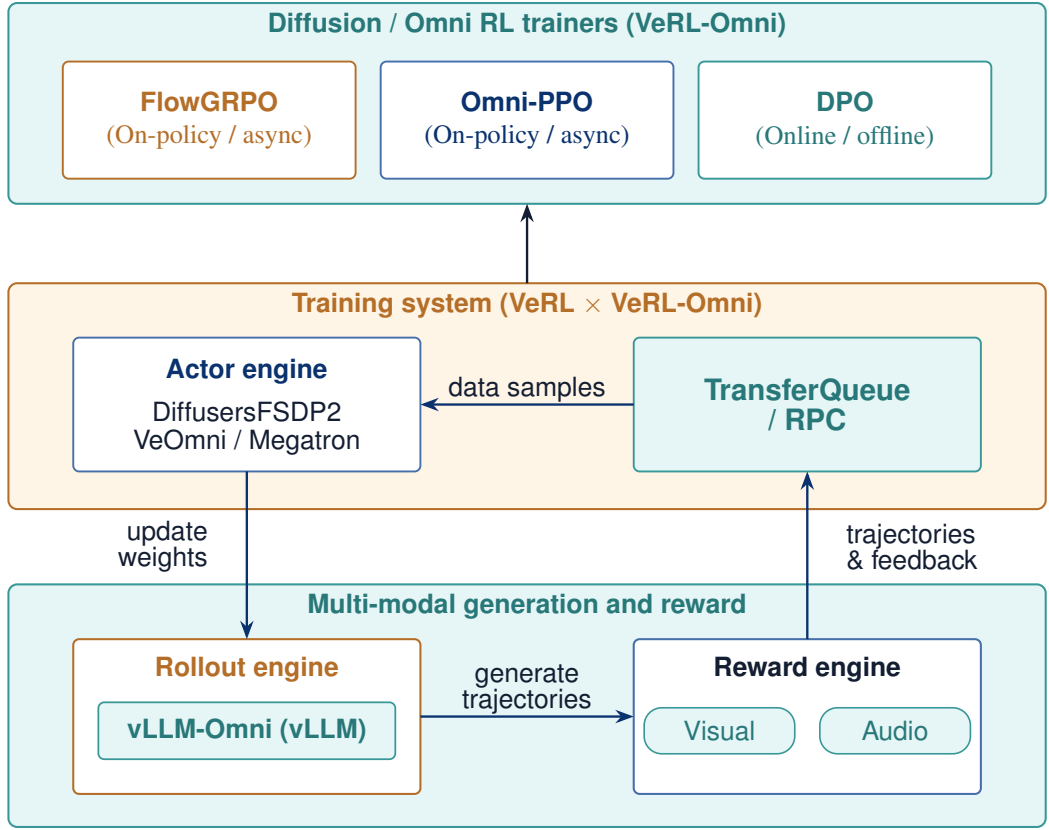

\paragraph{Training--Serving Handoff.}
Reinforcement-learning and post-training frameworks such as VeRL-Omni~\cite{verlomni_github} drive \sys{} as the rollout and serving backend, as shown in Figure~\ref{fig:verl_omni_arch}. 
The serving side exposes a narrow control surface for that handoff rather than owning the training loop.
Rollouts go through the asynchronous generate path or the OpenAI-compatible server.
Collective RPCs reach workers and stages for control operations; worker extensions can inject trainer-specific methods without restarting the process.
Diffusion LoRA add and remove APIs support adapter-weight handoff during training.
Sleep and wake free device memory between train and serve phases.
Custom pipeline re-initialization lets a stage swap into a logprob or other trainer-facing pipeline without tearing down the whole engine.
Weight synchronization and optimizer state remain outside \sys{}.
From the serving perspective, the contract is stable generation plus these control hooks.

\section{Supported Models and API Interfaces}
\label{sec:models-apis}

Section~\ref{sec:architecture} fixed the runtime mechanisms.
This section maps those mechanisms onto the external surfaces that applications actually call, and onto the model families that run as multi-stage pipelines in practice.
The APIs are intentionally thin adapters over the asynchronous online entrypoint and the orchestrator (Section~\ref{sec:arch-entrypoint-orch}).
They translate wire protocols into request admission, streaming progress, and session lifecycle events.
The model families are configured multi-stage pipelines over the engines in Section~\ref{sec:arch-engines}.
They reuse the same connector, KV, and session machinery rather than forking a server per modality.
Coverage evolves with releases; the living matrix lives in the project documentation.
Here we emphasize the stable contracts and representative topologies.

\subsection{API Interfaces}
\label{sec:apis}

\sys{} exposes two complementary serving surfaces.
OpenAI-compatible HTTP and WebSocket endpoints cover chat, speech, image, video, and realtime duplex traffic for multimodal assistants and media generation.
An OpenPI robot endpoint covers observation-to-action policy serving for embodied agents.
Both surfaces bind to one Omni process and one loaded multi-stage pipeline.
They differ in payload shape and session semantics, not in the underlying control plane.

\subsubsection{OpenAI-Compatible APIs}
\label{sec:apis-openai}

The OpenAI-compatible server is the default online entrypoint for product integrations.
By default, a single API server process hosts chat, speech, image, and video handlers alongside a small set of realtime WebSocket adapters.
For CPU-bound frontends, \texttt{--api-server-count} can start several API worker processes, each with its own frontend processor and orchestrator, over one shared set of stage engines.
This mode currently rejects diffusion stages.
Handlers validate protocol fields, construct Omni requests, and demultiplex orchestrator outputs into OpenAI-shaped responses.
Engine selection and stage topology remain behind the model registry and deploy configuration.
Operators therefore serve a Qwen-Omni pipeline, a CosyVoice3 TTS stack, or a Wan video DiT through the same binary, with different stage topologies and different active surfaces.

\paragraph{Modality Surfaces.}
Chat completions is the primary multimodal chat surface.
Clients send OpenAI-style messages, optionally with image, audio, or video content parts, and request text and/or audio modalities in the response.
When audio is not requested, the orchestrator can skip talker and waveform stages on Omni pipelines that declare that shortcut.
Streaming uses Server-Sent Events (SSE), including incremental text and base64 audio deltas for omni models.
The same chat surface can also drive selected diffusion image workloads through message content plus generation knobs in request extensions, which keeps simple T2I clients on a familiar path.

Speech synthesis follows the OpenAI speech contract, with optional batch and voice-management helpers.
Responses may be full audio bytes, chunked raw audio, or speech SSE events.
A streaming speech WebSocket accepts incremental text and flushes synthesis on an explicit end-of-input event, matching the streaming-input progress model in Section~\ref{sec:arch-entrypoint-orch}.
A separate audio-generate surface targets non-TTS diffusion audio generators such as Stable Audio, which share the speech product category but not the TTS stage topology.

Image generation and editing use the standard OpenAI image contracts.
These surfaces typically return base64 payloads or a file download body rather than token SSE.
Video serving offers both asynchronous jobs with status and content polling, and a synchronous path that returns raw video bytes.
Async jobs fit long DiT trajectories; the sync path fits short clips and tighter client loops.
Selected world-model and policy checkpoints also accept action-oriented fields on the video surface when the loaded pipeline exposes that mode.

\paragraph{Realtime Sessions.}
Longer interactive loops move to WebSockets.
A video-chat adapter streams camera frames and optional audio into an omni chat session, then returns text or audio deltas under the same request identity.
A realtime video adapter drives generated-video sessions through start, interaction, and stop events, which bind to the session layer in Section~\ref{sec:arch-session}.
An OpenAI-style realtime event stream covers speech agents; duplex mode routes the same traffic onto full-duplex session control instead of a one-shot completion API.
Across these sockets, the wire format changes; the orchestrator's notions of streaming updates, midway interaction, and session fencing stay the same.

Table~\ref{tab:api-surface} summarizes how the main surfaces map onto transports and typical stage topologies.
The table is a capability sketch rather than an endpoint catalog; concrete paths live in the serving documentation.

\begin{table}[!htbp]
\centering
\caption{OpenAI-compatible and robot serving surfaces in \sys{}.
Stage topologies are representative; deploy overlays may add or skip stages.}
\label{tab:api-surface}
\small
\setlength{\tabcolsep}{3.5pt}
\begin{tabular}{@{}llll@{}}
\toprule
\textbf{Surface} & \textbf{Transport} & \textbf{Typical stages} & \textbf{Streaming} \\
\midrule
Chat & HTTP/SSE & thinker$\to$talker$\to$code2wav & text/audio \\
Speech TTS & HTTP/SSE/WS & talker$\to$code2wav & chunked audio \\
Images & HTTP & DiT or AR$\to$DiT & full payload \\
Videos & HTTP job/sync & video DiT / flow & poll or bytes \\
Video realtime & WebSocket & omni AR or DiT session & frame/chunk \\
Duplex & WebSocket & llm$\to$tts$\to$code2wav & bidirectional \\
OpenPI robot & WS msgpack & policy DiT / AR--DiT & obs$\to$actions \\
\bottomrule
\end{tabular}
\end{table}

\subsubsection{Robot Serving}
\label{sec:apis-openpi}

Robot policy serving does not fit cleanly into chat completions.
Policies consume structured observations and emit action tensors on a tight control loop.
\sys{} therefore exposes a dedicated OpenPI-compatible WebSocket adapter.
The transport is msgpack rather than JSON.
After connect, the server advertises a policy configuration (action horizon, action keys, image resolution, and related metadata).
Each inference message carries a robot observation; the server returns actions.
The first call in a session may reset policy state, and an explicit reset clears continuity between episodes.
Idle sockets time out to bound leaked sessions.

This path is enabled only when the loaded model advertises a policy-server configuration.
Under the hood, observations enter a diffusion or AR--diffusion stage as structured extras, not as chat messages.
Session continuity, when required, uses the same session-oriented controls as world-model rollouts (Section~\ref{sec:arch-session}), rather than inventing a second robot runtime.
Product clients that already speak OpenPI can therefore target Omni without rewriting the observation--action loop into prompts.

\subsection{Supported Model Families}
\label{sec:models}

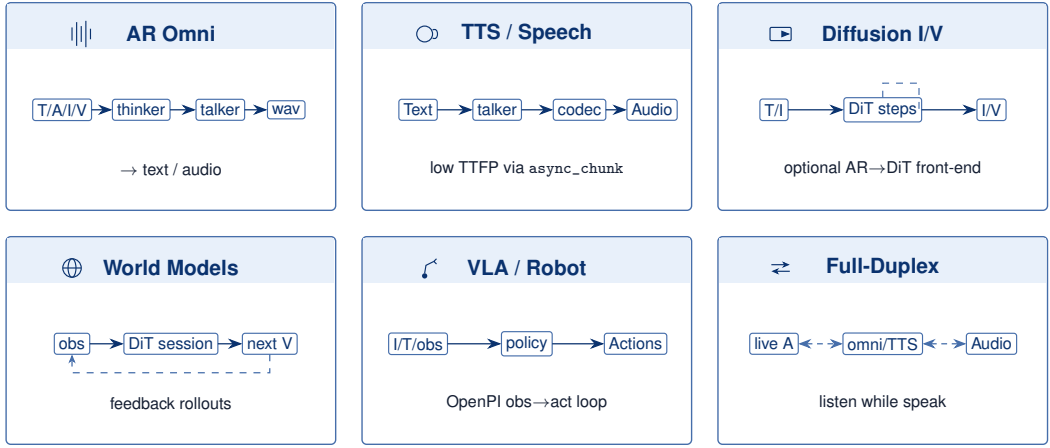
\begin{figure}[t]
  \centering
  \resizebox{\linewidth}{!}{
\begin{tikzpicture}[
  x=1cm, y=1cm,
  font=\sffamily,
  mini/.style={
    draw=omniBlueSoft, fill=white, rounded corners=1.2pt,
    line width=0.5pt, font=\sffamily\scriptsize, text=omniBlue,
    inner sep=2pt, align=center,
  },
]
  \foreach \x/\y/\name in {
    2.55/6.1/c1, 7.95/6.1/c2, 13.35/6.1/c3,
    2.55/2.55/c4, 7.95/2.55/c5, 13.35/2.55/c6} {
    \node[omni card, minimum width=5.0cm, minimum height=3.15cm]
      (\name) at (\x, \y) {};
  }

  \fill[omniBlueFill]
    ($(c1.north west)+(0.02,-0.02)$) rectangle ($(c1.north east)+(-0.02,-0.7)$);
  \node[font=\sffamily\bfseries\small, text=omniBlue] at (2.55, 7.2) {AR Omni};
  \begin{scope}[shift={(1.05,7.2)}, omniBlue, line width=0.5pt]
    \foreach \i/\h in {0/0.10,1/0.18,2/0.12,3/0.20,4/0.08}
      \draw (\i*0.07,-\h) -- (\i*0.07,\h);
  \end{scope}
  \node[mini] (o1) at (0.9, 6.05) {T/A/I/V};
  \node[mini] (o2) at (2.1, 6.05) {thinker};
  \node[mini] (o3) at (3.3, 6.05) {talker};
  \node[mini] (o4) at (4.3, 6.05) {wav};
  \draw[omni arrow, line width=0.55pt] (o1) -- (o2);
  \draw[omni arrow, line width=0.55pt] (o2) -- (o3);
  \draw[omni arrow, line width=0.55pt] (o3) -- (o4);
  \node[font=\sffamily\scriptsize, text=omniInk] at (2.55, 5.15) {$\rightarrow$ text / audio};

  \fill[omniBlueFill]
    ($(c2.north west)+(0.02,-0.02)$) rectangle ($(c2.north east)+(-0.02,-0.7)$);
  \node[font=\sffamily\bfseries\small, text=omniBlue] at (7.95, 7.2) {TTS / Speech};
  \begin{scope}[shift={(6.4,7.2)}, omniBlue, line width=0.5pt]
    \draw (0,0) circle[radius=0.12];
    \draw (0.12,0.06) .. controls (0.22,0.12) and (0.22,-0.12) .. (0.12,-0.06);
  \end{scope}
  \node[mini] (t1) at (6.3, 6.05) {Text};
  \node[mini] (t2) at (7.5, 6.05) {talker};
  \node[mini] (t3) at (8.75, 6.05) {codec};
  \node[mini] (t4) at (9.85, 6.05) {Audio};
  \draw[omni arrow, line width=0.55pt] (t1) -- (t2);
  \draw[omni arrow, line width=0.55pt] (t2) -- (t3);
  \draw[omni arrow, line width=0.55pt] (t3) -- (t4);
  \node[font=\sffamily\scriptsize, text=omniInk] at (7.95, 5.15) {low TTFP via \texttt{async\_chunk}};

  \fill[omniBlueFill]
    ($(c3.north west)+(0.02,-0.02)$) rectangle ($(c3.north east)+(-0.02,-0.7)$);
  \node[font=\sffamily\bfseries\small, text=omniBlue] at (13.35, 7.2) {Diffusion I/V};
  \begin{scope}[shift={(11.8,7.2)}, omniBlue, line width=0.5pt]
    \draw[rounded corners=0.6pt] (-0.16,-0.10) rectangle (0.16,0.10);
    \draw[fill=omniBlue] (0.02,-0.04) -- (0.12,0) -- (0.02,0.04) -- cycle;
  \end{scope}
  \node[mini] (d1) at (11.7, 6.05) {T/I};
  \node[mini] (d2) at (13.35, 6.05) {DiT steps};
  \node[mini] (d3) at (15.0, 6.05) {I/V};
  \draw[omni arrow, line width=0.55pt] (d1) -- (d2);
  \draw[omni arrow, line width=0.55pt] (d2) -- (d3);
  \draw[omniBlueSoft, dashed, line width=0.5pt]
    (d2.north) -- ++(0,0.22) -| ++(0.55,0) |- (d2.east);
  \node[font=\sffamily\scriptsize, text=omniInk] at (13.35, 5.15) {optional AR$\rightarrow$DiT front-end};

  \fill[omniBlueFill]
    ($(c4.north west)+(0.02,-0.02)$) rectangle ($(c4.north east)+(-0.02,-0.7)$);
  \node[font=\sffamily\bfseries\small, text=omniBlue] at (2.55, 3.65) {World Models};
  \begin{scope}[shift={(1.05,3.65)}, omniBlue, line width=0.5pt]
    \draw (0,0) circle[radius=0.14];
    \draw (0,0) ellipse[x radius=0.06, y radius=0.14];
    \draw (-0.14,0) -- (0.14,0);
  \end{scope}
  \node[mini] (w1) at (1.05, 2.5) {obs};
  \node[mini] (w2) at (2.55, 2.5) {DiT session};
  \node[mini] (w3) at (4.05, 2.5) {next V};
  \draw[omni arrow, line width=0.55pt] (w1) -- (w2);
  \draw[omni arrow, line width=0.55pt] (w2) -- (w3);
  \draw[omniBlueSoft, dashed, line width=0.55pt, -{Stealth[length=1.4mm,width=1.1mm]}]
    (w3.south) -- ++(0,-0.28) -| (w1.south);
  \node[font=\sffamily\scriptsize, text=omniInk] at (2.55, 1.6) {feedback rollouts};

  \fill[omniBlueFill]
    ($(c5.north west)+(0.02,-0.02)$) rectangle ($(c5.north east)+(-0.02,-0.7)$);
  \node[font=\sffamily\bfseries\small, text=omniBlue] at (7.95, 3.65) {VLA / Robot};
  \begin{scope}[shift={(6.4,3.65)}, omniBlue, line width=0.5pt]
    \fill (0,-0.12) circle[radius=0.04];
    \draw (0,-0.08) -- (0.02,0.04) -- (0.14,0.10);
    \draw (0.14,0.10) -- (0.18,0.06) (0.14,0.10) -- (0.18,0.14);
  \end{scope}
  \node[mini] (v1) at (6.3, 2.5) {I/T/obs};
  \node[mini] (v2) at (7.95, 2.5) {policy};
  \node[mini] (v3) at (9.6, 2.5) {Actions};
  \draw[omni arrow, line width=0.55pt] (v1) -- (v2);
  \draw[omni arrow, line width=0.55pt] (v2) -- (v3);
  \node[font=\sffamily\scriptsize, text=omniInk] at (7.95, 1.6) {OpenPI obs$\rightarrow$act loop};

  \fill[omniBlueFill]
    ($(c6.north west)+(0.02,-0.02)$) rectangle ($(c6.north east)+(-0.02,-0.7)$);
  \node[font=\sffamily\bfseries\small, text=omniBlue] at (13.35, 3.65) {Full-Duplex};
  \begin{scope}[shift={(11.8,3.65)}, omniBlue, line width=0.5pt]
    \draw[-{Stealth[length=1.2mm,width=1.0mm]}] (-0.14,0.05) -- (0.14,0.05);
    \draw[-{Stealth[length=1.2mm,width=1.0mm]}] (0.14,-0.05) -- (-0.14,-0.05);
  \end{scope}
  \node[mini] (f1) at (11.7, 2.5) {live A};
  \node[mini] (f2) at (13.35, 2.5) {omni/TTS};
  \node[mini] (f3) at (15.0, 2.5) {Audio};
  \draw[omni heavy bi, line width=0.55pt] (f1) -- (f2);
  \draw[omni heavy bi, line width=0.55pt] (f2) -- (f3);
  \node[font=\sffamily\scriptsize, text=omniInk] at (13.35, 1.6) {listen while speak};

  \omniBalanceLR
\end{tikzpicture}}
  \caption{Six workload families as multi-stage patterns on one runtime.}
  \label{fig:model-families}
\end{figure}

We group supported models by serving topology rather than by marketing category.
Families differ in which engines they instantiate and how stages are wired.
They share one orchestrator, one connector fabric, and one session abstraction.
Hybrid AR-plus-DiT pipelines appear inside the families that need them; they are not treated as a separate top-level category.
Near-identical checkpoint variants that share one architecture are collapsed into a single row.
Per-backend coverage lives in the project documentation and evolves with releases.
The subsections below expand the architecture hooks each family stresses.
Figure~\ref{fig:model-families} summarizes what each family does and the stage pattern it typically exercises.
Table~\ref{tab:supported-models} then lists supported models one row each, with family, default stage topology, input/output modalities, and primary serving surface.

{
\scriptsize
\setlength{\tabcolsep}{2.8pt}
\begin{longtable}{@{}p{3.05cm}p{1.15cm}p{2.85cm}p{1.85cm}p{1.4cm}p{1.65cm}@{}}
\caption{Supported models in \sys{}.
Each row is one model line.
Citations point to the original paper, arXiv preprint, or vendor technical report when available.
Stages are default deploy topologies; overlays may add, fuse, or skip stages.
Modalities: T\,=\,text, A\,=\,audio, I\,=\,image, V\,=\,video, Act\,=\,action/observation.}
\label{tab:supported-models}\\
\toprule
\textbf{Model} & \textbf{Family} & \textbf{Stages} & \textbf{Input} & \textbf{Output} & \textbf{Surface} \\
\midrule
\endfirsthead
\toprule
\textbf{Model} & \textbf{Family} & \textbf{Stages} & \textbf{Input} & \textbf{Output} & \textbf{Surface} \\
\midrule
\endhead
\midrule
\multicolumn{6}{r}{\emph{Continued on next page}} \\
\endfoot
\bottomrule
\endlastfoot
Qwen3-Omni~\cite{Qwen3-Omni} & AR Omni & thinker$\to$talker$\to$code2wav & T/A/I/V & T/A & Chat / realtime \\
Qwen2.5-Omni~\cite{Qwen2.5-Omni} & AR Omni & thinker$\to$talker$\to$code2wav & T/A/I/V & T/A & Chat \\
MiniCPM-o~4.5~\cite{minicpmo45} & AR Omni & llm$\to$tts$\to$code2wav & T/A/I/V & T/A & Chat / duplex \\
PersonaPlex~\cite{personaplex} & Duplex & Moshi-style S2S + session & A (+T) & A & Duplex \\
Nemotron VoiceChat~\cite{nemotronvoicechat} & Duplex & native S2S + session & A (+T) & A & Duplex \\
Ming-flash-omni~2.0~\cite{Ming-Flash-Omni} & AR Omni & thinker$\to$tts / thinker$\to$DiT & T/A/I & T/A/I & Chat / image \\
MiMo-Audio~\cite{mimoaudio} & AR Omni & fused thinker--talker$\to$code2wav & T/A & T/A & Chat \\
Covo-Audio-Chat~\cite{covoaudio} & AR Omni & omni audio chat stages & T/A & T/A & Chat \\
Dynin-Omni~\cite{dyninomni} & AR Omni & token2\{text,image,audio\} & T/A/I & T/A/I & Chat \\
Step-Audio2 & AR Omni & thinker$\to$token2wav & T/A & T/A & Chat \\
\midrule
Qwen3-TTS~\cite{Qwen3-TTS} & TTS & talker$\to$code2wav, or fused talker & T (+ref~A) & A & Speech \\
CosyVoice3~\cite{cosyvoice3} & TTS & talker$\to$code2wav (flow) & T (+ref~A) & A & Speech \\
Fish Speech S2 Pro~\cite{fishspeech} & TTS & slow AR$\to$DAC codec & T (+ref~A) & A & Speech \\
Voxtral TTS~\cite{voxtral} & TTS & audio gen.$\to$tokenizer & T & A & Speech \\
Higgs-Audio v2~\cite{higgsaudio} & TTS & talker$\to$code2wav & T (+ref~A) & A & Speech \\
Higgs-Audio v3~\cite{higgsaudio} & TTS & talker$\to$code2wav & T (+ref~A) & A & Speech \\
IndexTTS-2~\cite{indextts2} & TTS & talker$\to$S2Mel & T (+ref~A) & A & Speech \\
IndexTTS-2.5~\cite{indextts2} & TTS & talker$\to$S2Mel & T (+ref~A) & A & Speech \\
MOSS-TTS / TTSD~\cite{mosstts} & TTS & talker$\to$codec & T (+ref~A) & A & Speech \\
MOSS-TTS-Realtime~\cite{mosstts} & TTS & talker$\to$codec & T (+ref~A) & A & Speech \\
MOSS-TTS-Local~\cite{mosstts} & TTS & talker$\to$codec & T (+ref~A) & A & Speech \\
MOSS-TTS-Nano~\cite{mosstts} & TTS & compact TTS stages & T & A & Speech \\
MOSS-VoiceGenerator~\cite{mosstts} & TTS & talker$\to$codec & T & A & Speech \\
GLM-TTS~\cite{glmtts} & TTS & talker$\to$vocoder & T (+ref~A) & A & Speech \\
Ming-omni-tts~\cite{Ming-Omni} & TTS & LLM$\to$audio-VAE & T (+ref~A) & A & Speech \\
Audex & TTS & thinker$\to$code2wav & T & A & Speech \\
dots.tts & TTS & single AR talker & T & A & Speech \\
Gepard-1.0 & TTS & single AR codec TTS & T (+ref~A) & A & Speech \\
Audio8 TTS & TTS & slow AR$\to$codec & T & A & Speech \\
OmniVoice~\cite{omnivoice} & TTS & single DiT TTS & T (+ref~A) & A & Speech \\
VoxCPM2~\cite{voxcpm2} & TTS & latent generator & T (+ref~A) & A & Speech \\
Stable-Audio-Open~\cite{stableaudio} & Audio & single DiT audio & T & A & Audio gen. \\
MiniMax-Music3 & Audio & talker$\to$decoder & T & A & Audio gen. \\
MOSS-SoundEffect~\cite{mosstts} & Audio & talker$\to$codec & T & A & Audio gen. \\
\midrule
Qwen-Image~\cite{Qwen-Image} & Image & single DiT & T & I & Image \\
Qwen-Image-Edit~\cite{Qwen-Image} & Image & single DiT (edit) & T/I & I & Image edit \\
Qwen-Image-Layered~\cite{Qwen-Image} & Image & single DiT & T/I & I & Image \\
FLUX.1-dev / schnell~\cite{FLUX.1} & Image & single DiT & T & I & Image \\
FLUX.1-Kontext~\cite{FLUX.1} & Image & single DiT (edit) & T/I & I & Image edit \\
FLUX.2-dev~\cite{FLUX.1} & Image & single DiT & T & I & Image \\
FLUX.2-klein~\cite{FLUX.1} & Image & single DiT & T & I & Image \\
Z-Image-Turbo~\cite{zimage} & Image & single DiT & T & I & Image \\
SDXL~\cite{sdxl} & Image & single DiT & T & I & Image \\
SD3.5~\cite{sd3} & Image & single DiT & T & I & Image \\
LongCat-Image~\cite{longcatimage} & Image & single DiT & T & I & Image \\
LongCat-Image-Edit~\cite{longcatimage} & Image & single DiT (edit) & T/I & I & Image edit \\
Ovis-Image~\cite{ovisimage} & Image & single DiT & T & I & Image \\
ERNIE-Image~\cite{ernieimage} & Image & single DiT & T & I & Image \\
OmniGen2~\cite{omnigen2} & Image & single DiT & T/I & I & Image \\
HiDream-I1~\cite{hidream} & Image & single DiT & T & I & Image \\
Krea~2~\cite{krea2} & Image & single DiT & T & I & Image \\
Boogu-Image~\cite{booguimage} & Image & single DiT & T/I & I & Image \\
NextStep-1.1~\cite{nextstep11} & Image & single DiT & T & I & Image \\
GLM-Image~\cite{GLMImage2026} & Image & AR$\to$DiT & T & I & Image \\
HunyuanImage-3.0~\cite{hunyuanimage} & Image & AR$\to$DiT / DiT-only & T & I & Image \\
BAGEL~\cite{bagel} & Image & thinker$\to$DiT / DiT-only & T/I & T/I & Image / chat \\
MammothModa2~\cite{mammothmoda2} & Image & AR$\to$DiT & T/I & T/I & Image / chat \\
SenseNova-U1~\cite{sensenovau1} & Image & DiT-only & T/I & I & Image \\
\midrule
Wan2.1/2.2-T2V~\cite{wan} & Video & single DiT & T & V & Video \\
Wan2.2-I2V~\cite{wan} & Video & single DiT & T/I & V & Video \\
Wan2.2-S2V~\cite{wan} & Video & single DiT & T/A & V & Video \\
Wan2.1/2.2-VACE~\cite{wan} & Video & single DiT & T/I/V & V & Video \\
LTX-2 / LTX-2.3 / LTX-2.5~\cite{ltx2} & Video & one-stage DiT & T/I & V & Video \\
LTX-2 distilled~\cite{ltx2} & Video & two-stage DiT & T/I & V & Video \\
SANA-Video~\cite{sanavideo} & Video & single DiT (T2V/I2V) & T/I & V & Video \\
HunyuanVideo-1.5~\cite{hunyuanvideo15} & Video & single DiT (T2V/I2V) & T/I & V & Video \\
Helios~\cite{heliosvideo} & Video & DiT / pyramid & T/I & V & Video \\
LingBot-Video~\cite{lingbotvideo} & Video & dense / MoE DiT & T & V & Video \\
MiniMax~H3~\cite{minimaxh3} & Video & joint DiT (T2VA/Ref2VA) & T/I/V/A & V/A & Video \\
MAGI-2 & Video & single DiT (T2VA/I2VA) & T/I & V/A & Video \\
LongCat-Video-Avatar-1.5 & Video & single DiT (AT2V/AI2V) & T/A/I & V & Video \\
Lance~\cite{lancevideo} & Video & DiT video & T/I & V & Video \\
\midrule
Cosmos3-Nano/Super~\cite{cosmos} & World & DiT omni (+A/Act) & T/I/V/Act & I/V/A/Act & Video / OpenPI \\
DreamZero-DROID~\cite{dreamzero} & World & AR--DiT + session & I/Act & V/Act & Video / OpenPI \\
LingBot-World~2.0 & World & single DiT session & T/I & V & Video realtime \\
SANA-WM & World & single DiT (I2V + camera) & T/I/Act & V & Video \\
\midrule
GR00T~N1.7~\cite{grootn1} & VLA & policy DiT & I/T/Act & Act & OpenPI \\
$\pi_0$~\cite{pi0} & VLA & policy DiT & I/T/Act & Act & OpenPI \\
InternVLA-A1~\cite{internvlaa1} & VLA & VLA pipeline & I/T/Act & Act & Offline / OpenPI \\
\end{longtable}
}

\subsubsection{Autoregressive Omni}
\label{sec:models-omni}

Autoregressive omni models are the canonical multi-stage AR pipeline, typified by Qwen2.5/3-Omni (thinker$\to$talker$\to$code2wav) and MiniCPM-o (llm$\to$tts$\to$code2wav)~\cite{Qwen2.5-Omni,Qwen3-Omni}.
The thinker (or LLM) stage performs multimodal understanding and text generation; the talker emits speech codecs; and code2wav reconstructs waveforms.
Ming-flash-omni and related lines add modality branches such as thinker$\to$DiT for image generation~\cite{Ming-Omni,Ming-Flash-Omni}.
Other omni stacks in tree fuse thinker and talker, or insert ASR front-ends, but they still fit the same multi-stage pattern.
Beyond those default topologies, production Omni serving leans on three architecture hooks already defined in Section~\ref{sec:architecture}.

Omni speech paths enable \texttt{async\_chunk} on talker-to-code2wav and similar edges so waveform decoding overlaps codec prediction (Section~\ref{sec:arch-entrypoint-orch}).
In the conventional batch handoff, the waveform stage would wait until the talker finished an entire utterance, which serializes codec prediction and vocoding and inflates time-to-first-packet (TTFP).
Under \texttt{async\_chunk}, the orchestrator pre-warms the downstream stage, workers publish codec chunks on the connector as soon as they are produced, and stage-specific chunk processors assemble those chunks into runnable decoder inputs.
The orchestrator then suppresses the redundant control-plane forward that the batch path would otherwise issue for the same edge.
Thinker and talker decoding continue while early waveform packets already leave toward the client.
Deploy overlays choose which edges stream and which remain full-payload, so text-only or non-speech branches need not pay the chunk protocol cost.

The same loaded Omni graph often serves mixed traffic: some requests want spoken replies, others want text understanding only.
When the client requests text-only modalities, the orchestrator skips talker and code2wav stages if the pipeline declares that shortcut.
Skipped stages do not consume scheduling slots or connector bandwidth, even though their weights may remain resident for concurrent speech requests.
This selectivity is expressed as orchestrator progress rules rather than as a second model server, which keeps chat completions and omni speech on one process while still avoiding unnecessary speech compute.

Thinker, talker, and waveform stages have mismatched memory and latency profiles.
Multimodal thinkers are weight-heavy and benefit from large-batch AR scheduling; vocoders are often lighter in batch dimension but sensitive to placement relative to the audio path.
\sys{} therefore allows these stages to sit on different devices or hosts, with independent replica pools and load-balancing policies (Section~\ref{sec:arch-engines}).
Sticky routing pins a chunked request to the same downstream replica so that \texttt{async\_chunk} state remains coherent across packets.
Operators can scale understanding capacity without over-provisioning vocoders, which is difficult when the entire Omni stack is fused into one monolithic process.
Qwen3-Omni also exposes a turn-based \texttt{/v1/realtime} path that attaches continuous input to a live request identity, without resume, playback acknowledgement, or overlap handling.
That path is a realtime chat API, not a full-duplex session.

\subsubsection{Speech Models}
\label{sec:models-tts}

Text-to-speech (TTS) and broader speech-generation models drop the full omni
understanding stack and keep a talker or codec predictor plus a vocoder or
flow-matching decoder, as in Qwen3-TTS and CosyVoice3~\cite{cosyvoice3}.
Other lines include Fish-Speech-style AR-plus-codec stacks, Voxtral-style generation pipelines, and single-stage diffusion TTS.
For this family, TTFP and stable streaming under interactive input dominate the systems agenda more than peak tokens-per-second on a thinker.

Deploy defaults for CosyVoice3 and Qwen3-TTS turn on \texttt{async\_chunk} so codec frames enter the vocoder while the AR talker is still decoding.
That overlap is the same inter-stage mechanism used by Omni speech, but TTS deployments lean on it even more aggressively because there is no heavy thinker stage to hide behind.
On the wire, chunked PCM, speech SSE events, and the incremental-text speech WebSocket all demultiplex through the orchestrator's streaming output path rather than through a TTS-specific runtime.
The result is that online speech serving reuses Omni's control and data planes end to end: stage advancement, connector chunk publish/subscribe, and client delivery share one request identity.
Operators who need coarser handoffs can disable async chunking per deploy overlay without changing the API surface.
On Qwen3-TTS, an adaptive chunk ramp can widen or shrink chunk watermarks from vocoder buffer feedback so TTFP and underrun risk are not locked to a single static size.
Qwen3-TTS also offers an opt-in fused single-stage pipeline for one GPU, where the talker owns a stateful streaming codec decoder and returns PCM directly.
This removes the code2wav stage and its connector hop.
Section~\ref{sec:eval-tts} compares it with the two-stage default.

The AR talker is where continuous batching, compilation, and prefix reuse of prompt or reference embeddings pay off first.
Vocoders and flow-matching decoders are often memory-heavy, less amenable to large-batch AR tricks, and sometimes prefer different quantization or compile settings.
Disaggregating talker and vocoder therefore lets each stage take a different efficiency profile from Section~\ref{sec:platforms-efficiency}: quantize and graph-capture the talker early, keep the vocoder at higher precision until validated, and size replica counts independently.
When the codec decoder is light enough to share a GPU with the talker, fusing the two removes the inter-stage hop, and the fused Qwen3-TTS pipeline on one GPU delivers $1.2$--$2.2\times$ the audio throughput of the two-stage default on two GPUs (Table~\ref{tab:tts}).
When both stages remain co-located, the same pipeline still runs; only the binding of workers to devices changes.
This separation is what makes TTS look like a narrow Omni pipeline rather than a one-off speech binary.

Production TTS rarely synthesizes from bare text alone.
Reference audio, language tags, speaker or task-type fields, and voice-design knobs must reach the talker stage without inventing a second control channel.
\sys{} folds those fields into the speech request and stage configuration so that voice cloning and multi-speaker serving stay inside the Omni engine.
Interactive clients further need streaming input: text arrives in segments, synthesis should begin early, and an explicit end-of-input event should flush the remaining audio.
The entrypoint models that pattern as resumable stage~0 updates plus a final flush, which matches speech WebSocket UIs and avoids the latency of waiting for a complete HTTP body before any codec token is produced.

\subsubsection{Diffusion Models}
\label{sec:models-dit}

Diffusion image and video models occupy DiT and flow-matching engines, from single-stage Qwen-Image, FLUX, and Wan pipelines to hybrid AR$\to$DiT systems such as GLM-Image, HunyuanImage, and BAGEL~\cite{Qwen-Image,FLUX.1,wan,GLMImage2026,hunyuanimage,bagel,DiT}.
MiniMax~H3 is a joint video--audio DiT on the same surface: one stage serves T2VA and reference-conditioned Ref2VA without a second orchestrator~\cite{minimaxh3}.
Some large video pipelines further disaggregate a text-encoder stage from the DiT body when conditioning is heavy enough to warrant its own placement.
Video trajectories amplify both step count and sequence length, so specialization concentrates on step reuse, parallel execution, and memory movement inside the diffusion stage (Section~\ref{sec:platforms-efficiency}).
The orchestrator still sees an ordinary stage; the optimizations attach as engine-local options rather than as alternate runtimes.

Iterative denoisers repeatedly evaluate similar transformer blocks across timesteps.
TeaCache estimates when consecutive steps are similar enough to reuse modulated activations; Cache-DiT provides a complementary backend for step-to-step reuse on supported models.
Either backend reduces denoise FLOPs on image and long video jobs without changing stage topology, connector edges, or API handlers.
They are alternatives rather than a stack: enabling both is unsupported, and step-execution debugging modes disable caches so that numerical comparisons stay meaningful.
Secondary caches exist in tree for selected models, but TeaCache and Cache-DiT are the first-class paths operators should prefer.
Because caching is stage-local, a hybrid AR$\to$DiT pipeline can cache the DiT stage while the AR front-end continues to use ordinary AR scheduling.

High-resolution images and long videos exceed single-GPU activation memory even when weights fit.
Ulysses and Ring sequence parallelism shard the sequence dimension across devices with different communication patterns; CFG parallelism evaluates conditional and unconditional branches concurrently; tensor, pipeline, and HSDP shard or replicate weights; expert parallel targets MoE DiTs; and VAE patch parallel reduces peak memory during decode.
These strategies are selected per diffusion stage and compose with TeaCache or Cache-DiT only under documented compatibility constraints.
Importantly, parallelism does not require a second orchestrator: replica and device meshes remain bindings on the same stage node that chat or TTS stages use for AR workers.
Operators therefore scale a Wan or HunyuanVideo stage the same way they scale any other Omni stage, while choosing diffusion-specific parallel knobs inside that stage's engine config.

When parallelism alone is insufficient, module-wise or layerwise CPU offload moves inactive DiT blocks or encoders off device between uses, which is especially useful for large video models and for train--serve colocation.
Regional compilation lowers kernel-launch overhead on hot denoise blocks without demanding a single static full graph for every resolution and step count.
Hybrid image systems add a further cross-stage concern: an AR or multimodal thinker must condition the DiT.
Stage-level KV transfer carries packed cache state along the AR$\to$DiT edge so that the diffusion stage need not rebuild conditioning from a textual detour alone, and diffusion-side paged KV can receive those blocks under scheduler ownership when the pipeline enables it
(Section~\ref{sec:arch-kv}).
Taken together, caches, parallelism, offload, compile, and KV handoff are how diffusion families specialize inside the shared runtime.

\subsubsection{World Models}
\label{sec:models-world}

World foundation models such as Cosmos3 and DreamZero predict future visual states under text, past observations, and sometimes actions~\cite{cosmos}.
Unlike one-shot image generation, a rollout is a sequence of interactions in which yesterday's prediction becomes today's conditioning.
The pipeline may still be a single diffusion or AR--diffusion stage; what changes is that the control plane must keep session-scoped state alive across those interactions (Section~\ref{sec:arch-session}).

Attention KV is not a sufficient carrier for world-model continuity.
Latents, feature buffers, and other non-KV tensors must survive from one forward to the next under an explicit session identity.
\sys{} retains that state in the session state manager, with byte accounting and host offload when device memory is tight, so long rollouts do not silently accumulate unbounded GPU residency.
Previously generated frames can be fed back as observations without rebuilding the entire engine request from scratch.
Eviction and stop events reclaim state when a session ends, which keeps world-model serving from leaking memory across clients.
This is the main reason world-model serving differs from Wan-style one-shot video even when both use a DiT engine.

Cosmos3 consolidates text-to-image, text-to-video, image-to-video, video-to-video, optional sound, and action-oriented modes behind one pipeline configuration.
Mode selection is expressed as request modalities and generation arguments on a shared engine, not as a process-per-mode deployment.
That design matters operationally: the same weights, connector settings, and efficiency knobs serve several product surfaces, while the session layer decides whether a call is a cold generation or a continuation.
DreamZero-style AR--diffusion engines similarly keep a single stage warm and vary the interaction payload across steps.
Realtime world-model serving can further use tick-style AR--diffusion sessions that advance the warm stage on a fixed interaction cadence without rebuilding the pipeline between frames.

Some world-model checkpoints are also policy-capable.
When a Cosmos or DreamZero deploy advertises policy-server metadata, the same session-backed engine can speak OpenPI for action chunks while still serving video rollouts on the generative APIs.
The optimization is boundary reuse: observation streams, session reset, and latent continuity do not need a parallel robot stack beside the world-model server.
Deployments that only need generative video simply leave the policy surface disabled.

\subsubsection{VLA and Robot Policy}
\label{sec:models-vla}

Vision--language--action models map perception and language to robot actions~\cite{openvla,grootn1}.
GR00T N1.7, $\pi_0$, InternVLA-A1, and policy variants of DreamZero or Cosmos run as diffusion or AR--diffusion stages.
Their serving contract is an observation-to-action loop, so the important specializations are protocol boundary, concurrency policy, and episode lifecycle rather than speech-style TTFP.

Chat completions are a poor fit for closed-loop control: packing camera tensors into multimodal chat messages adds framing overhead and invites prompt-shaped APIs where a policy server is expected.
The OpenPI adapter instead exchanges msgpack observations and action tensors over a dedicated WebSocket, and the diffusion policy stage consumes structured robot observations as engine extras.
Policy configuration (action horizon, action keys, image resolution, and related metadata) is advertised at connect time so that clients can validate embodiment assumptions before the first infer call.
This keeps actuation on a stable binary contract while still running inside the Omni multi-stage runtime.

Robot policies often require deterministic per-client state across an action horizon and are rarely throughput-bound in the LLM sense.
Deployments therefore commonly admit one sequence at a time and rely on sticky session affinity so that successive observations land on the same worker that holds the warm policy state.
Continuous batching across many unrelated robots is less valuable than avoiding cross-talk between episodes and keeping inference latency predictable for the control loop.
In that regime, stage replicas still help with multi-robot fan-out, but load balancing prefers affinity over least-queue aggression.

Explicit reset clears continuity between episodes without tearing down device contexts or reloading weights.
That call is the robot analogue of ending a world-model session: latent or policy state from the previous episode must not leak into the next.
Idle timeouts close abandoned sockets so that leaked robot clients cannot pin GPU memory indefinitely.
Together, reset and timeout give operators a small control surface for long-lived physical interaction while sharing the same engine process used for offline VLA evaluation.

\subsubsection{Full-Duplex}
\label{sec:models-duplex}

Full-duplex agents keep listening while speaking.
MiniCPM-o duplex configurations, PersonaPlex, Nemotron VoiceChat, and related realtime paths reuse omni or TTS pipelines under session mode; they do not introduce a separate duplex engine family~\cite{personaplex,nemotronvoicechat}.
The specializations are therefore session-level rather than model-architectural.

A duplex turn is not a strict request/response pair.
Microphone audio or text may arrive while waveform stages are still emitting chunks for the previous partial reply.
\sys{} keeps both directions under one session identity so that streaming updates and streaming outputs do not fork into disconnected control planes (Section~\ref{sec:arch-session}).
Speech edges continue to use \texttt{async\_chunk} where configured, which means duplex inherits Omni's low-TTFP path instead of inventing a second chunk protocol.
Video-chat style inputs extend the same idea to visual frames that condition an ongoing omni request.

Turn taking creates correctness hazards that one-shot chat never sees: stale audio from a previous turn must not be treated as context for the next, and too many concurrent duplex sessions can exhaust GPU memory.
Turn and epoch fences drop or isolate stale segments at session boundaries, and admission control rejects new sessions once a deploy-configurable concurrent-session cap is reached.
Native barge-in of an in-flight reply is model-specific: MiniCPM-o~4.5 advertises it, while PersonaPlex and Nemotron VoiceChat do not.
These controls sit beside the ordinary Omni multi-stage pipeline rather than inside each model implementation, so several duplex-capable checkpoints can share one control plane.
Only the native \texttt{/v1/duplex} protocol is still documented as experimental.
The API therefore presents a realtime socket, while the runtime remains the AR omni pipeline held open for the lifetime of a conversation.

\section{Evaluation}
\label{sec:eval}

We evaluate \sys{} as a serving system on NVIDIA H100 and H200 GPUs.
The speech tables (Qwen3-Omni Fixed-Len, Random-MM, and text-only; the TTS
portfolio; MiniCPM-o) are measured on H200, comparing
each model's default deployment with its opt-in optimized profile.
The image, video, and world-model tables come from the public H100 multimodal
nightly CI at freeze day \textbf{2026-09-14}, or the H100 recipe baseline that
feeds that night when a history row is missing.
The Qwen3-Omni async-chunk and replica tables use H100 recipe baselines averaged
over the nightlies of 2026-07-05 to 2026-07-11, because those sweeps no longer
run every night.
The campaign covers Qwen3-Omni speech serving with async-chunk and replica
sweeps, a TTS portfolio, diffusion image and video recipes, Cosmos3/MiniMax
video cells, and MiniCPM-o duplex health metrics.
Cross-framework overlays and engineering-health aggregates are left out of
this revision.

\subsection{Experimental Setup}
\label{sec:eval-setup}

\stitle{Hardware and software}
Cells run on NVIDIA H100 GPUs in the nightly multimodal pool, except the TTS
and MiniCPM-o tables, which run on H200 GPUs.
GPU counts follow each recipe (Omni 2--3, TTS/Voxtral 2 by default and 1 for
the fused Qwen3-TTS profile, Qwen-Image 1 or 4,
Wan 1, Cosmos3 demo 2, MiniMax-H3 4).
Software SHAs are those recorded by the nightly job of each table's night.

\stitle{Recipes and load model}
Tables cite recipes under \texttt{tests/dfx/perf/tests/*.json}.
Closed-loop concurrency~$C$ keeps up to $C$ requests in flight; open-loop
Random-MM uses an offered \texttt{request\_rate}.
Omni Fixed-Len and TTS use $C\in\{1,4,8,16,32\}$ (TTS also $C{=}64$);
diffusion is $C{=}1$ except the Qwen-Image step-execution sweep
($C\in\{1,2,4,8\}$).

\stitle{Workloads}
Fixed-Len-2500/900 stresses 2{,}500/900 text tokens with text+audio outputs
(async-chunk vs.\ no-async-chunk).
Random-MM injects synthetic audio or image+video at
\texttt{request\_rate}${}\in\{0.1,0.5,1.0\}$.
A text-only Omni arm is compared to thinker-only vLLM; a 3-GPU cell doubles
talker/code2wav replicas.
TTS uses Seed-TTS English texts.
Image covers $512^{2}$/20-step and $1536^{2}$/35-step stills plus Cache-DiT
under Ulysses2+CFG2.
Wan2.2-I2V-A14B uses $832{\times}480$, 81 frames, 4 steps (plus USP2 and
$1280{\times}720$/121 frames); Cosmos3-Nano follows the official demo;
MiniMax-H3 uses $1344{\times}768$, 209 frames, 8 steps.
MiniCPM-o~4.5 reports challenge rows at $C\in\{1,4,8\}$ and one duplex
Seed-TTS row at $C{=}1$.

\stitle{Metrics}
TTFT is the delay until the first text token leaves the thinker.
TTFP is the delay until the first audio packet; it is what a listener
experiences as time-to-first-sound on speech outputs.
TPOT is the mean interval between subsequent text tokens (where the recipe
exports it).
E2EL is wall-clock time from client submission until the response completes.
RTF is the ratio of audio synthesis time to produced audio duration;
RTF${}<{}1$ means generation finishes faster than real-time playback.
TTS tables report median audio TTFP, median RTF, and audio-seconds per second
because the endpoint streams audio chunks rather than a text-token timeline.
Diffusion tables add mean latency, req/s, and peak GPU memory when the CI
export includes them.
All reported cells complete successfully.

\subsection{Qwen3-Omni Serving}
\label{sec:eval-omni}

\stitle{Concurrency scaling}
Qwen3-Omni is a three-stage thinker$\to$talker$\to$code2wav pipeline.
Under the default async-chunk route, downstream stages can start consuming
partial talker output instead of waiting for the full codec sequence, which
keeps the audio path overlapped with ongoing thinker decoding.
Table~\ref{tab:omni-fixedlen} sweeps Fixed-Len-2500/900 on two H200 GPUs under the default V1 deployment and the opt-in MRv2 profile.
Mean TTFT and E2EL grow with concurrency because more in-flight requests queue on the shared thinker and audio stages, yet mean RTF stays below one through $C{=}32$ in both deployments.
MRv2 roughly halves mean E2EL and RTF at $C{\ge}16$ (75.1 to 39.5\,s and 0.42 to 0.21 at $C{=}32$) and cuts mean audio TTFP from 1065 to 500\,ms at $C{=}32$, while thinker TTFT is slightly higher (336 vs.\ 322\,ms at $C{=}32$).
On the measured commit, the MRv2 profile fails at startup because shared MRv2 model-state code reads a streaming-decoder attribute that only the fused Qwen3-TTS talker defines.
All MRv2 Qwen3-Omni numbers use a one-line guard for that attribute and are otherwise unmodified.

\begin{figure}[t]
  \centering
  \includegraphics[width=\linewidth]{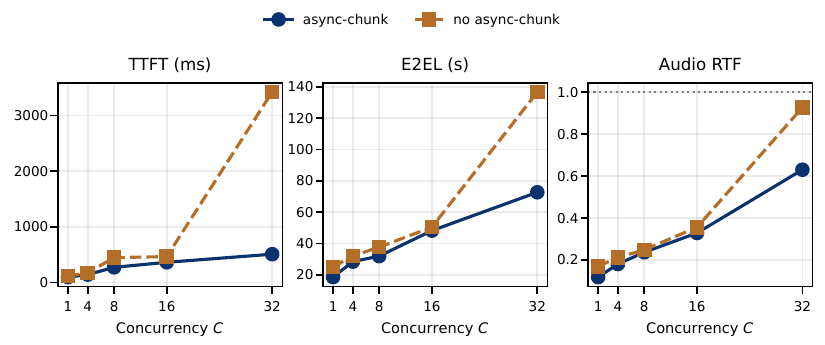}
  \caption{Qwen3-Omni Fixed-Len-2500/900 on $2{\times}$H100: async-chunk
  versus no-async-chunk for mean TTFT, E2EL, and audio RTF.}
  \label{fig:omni-fixedlen}
\end{figure}
\begin{table}[!htbp]
  \centering
  \small
  \caption{Qwen3-Omni Fixed-Len-2500/900 concurrency sweep (async-chunk, two
  NVIDIA H200, commit \texttt{ee8fdab1d}). ``Default'' is the nightly V1 recipe;
  ``MRv2'' is the H200 MRv2 profile from PR~\#8223 with devices mapped to the
  same two GPUs, run with a one-line startup fix described in the text. Same warm protocol as Table~\ref{tab:tts};
  medians over three repetitions.}
  \label{tab:omni-fixedlen}
  \begin{tabular}{r rr rr rr rr}
    \toprule
    & \multicolumn{2}{c}{TTFT mean (ms)} & \multicolumn{2}{c}{TTFP mean (ms)} & \multicolumn{2}{c}{E2EL mean (s)} & \multicolumn{2}{c}{RTF mean} \\
    \cmidrule(lr){2-3}\cmidrule(lr){4-5}\cmidrule(lr){6-7}\cmidrule(lr){8-9}
    $C$ & default & MRv2 & default & MRv2 & default & MRv2 & default & MRv2 \\
    \midrule
    1  & 96.0  & 110.7 & 241.6  & 212.1 & 15.04 & 12.34 & 0.109 & 0.083 \\
    4  & 118.0 & 140.3 & 319.9  & 256.2 & 27.24 & 19.61 & 0.132 & 0.090 \\
    8  & 156.4 & 206.9 & 439.5  & 355.8 & 32.89 & 19.20 & 0.178 & 0.111 \\
    16 & 212.2 & 242.1 & 687.4  & 379.6 & 48.02 & 26.91 & 0.261 & 0.147 \\
    32 & 322.2 & 336.3 & 1064.6 & 500.4 & 75.06 & 39.53 & 0.423 & 0.212 \\
    \bottomrule
  \end{tabular}
\end{table}

\stitle{Random-MM modality arms}
Random-MM holds the offered \texttt{request\_rate} fixed while varying input
modality and whether the client requests text+audio or text-only output
(Table~\ref{tab:omni-random-mm}).
Text+audio rows pay the full talker and code2wav path, so E2EL reflects both
language decoding and streaming synthesis.
On those rows MRv2 lowers mean E2EL by 15--33\% (5.28 to 3.53\,s on the
audio+image+video arm) and mean RTF from 0.10--0.15 to 0.09--0.11, with
thinker TTFT within 3\% of the default.
Text-only arms skip downstream speech stages and finish in under one second
of E2EL under both deployments.
Rows are not directly comparable as a quality ranking: input token load and
decoding work differ across modality arms.

\begin{table}[!htbp]
  \centering
  \small
  \caption{Qwen3-Omni Random-MM under open-loop request rates (async-chunk, two
  NVIDIA H200, commit \texttt{ee8fdab1d}). ``Default'' is the nightly V1
  recipe; ``MRv2'' is the PR~\#8223 profile as in Table~\ref{tab:omni-fixedlen}.
  Medians over three repetitions.}
  \label{tab:omni-random-mm}
  \begin{tabular}{l l r rr rr rr}
    \toprule
    & & & \multicolumn{2}{c}{TTFT mean (ms)} & \multicolumn{2}{c}{E2EL mean (s)} & \multicolumn{2}{c}{RTF mean} \\
    \cmidrule(lr){4-5}\cmidrule(lr){6-7}\cmidrule(lr){8-9}
    Input MM & Out & Rate & default & MRv2 & default & MRv2 & default & MRv2 \\
    \midrule
    audio             & text+audio & 0.1 & 271.8 & 278.0 & 3.11 & 2.64 & 0.101 & 0.085 \\
    image+video       & text+audio & 0.5 & 198.5 & 201.9 & 3.83 & 2.83 & 0.117 & 0.089 \\
    audio+image+video & text+audio & 1.0 & 396.9 & 403.8 & 5.28 & 3.53 & 0.154 & 0.107 \\
    audio             & text       & 0.1 & 318.4 & 262.1 & 0.72 & 0.66 & --- & --- \\
    image+video       & text       & 0.5 & 195.9 & 199.4 & 0.63 & 0.63 & --- & --- \\
    audio+image+video & text       & 1.0 & 372.4 & 388.8 & 0.89 & 0.92 & --- & --- \\
    \bottomrule
  \end{tabular}
\end{table}

\stitle{Text-only route}
Table~\ref{tab:omni-text} compares the Omni text-only route with a thinker-only vLLM endpoint on the same Fixed-Len lengths, with both running the thinker on a single H200.
At $C{=}1$, the two are within 6\% on TTFT and E2EL.
At $C{=}32$, mean E2EL is identical at 13.7\,s, but Omni TTFT is 860\,ms versus 562\,ms; we have not isolated how much of that gap comes from the stage graph and orchestrator in front of the thinker.
The MRv2 profile is slower on this route, averaging about 1.0\,s TTFT and 16.5\,s E2EL at $C{=}32$.
During two separate runs, its thinker exited in the third $C{=}32$ repetition and failed 32 of 128 requests, so the table reports the default deployment.

\begin{table}[!htbp]
  \centering
  \small
  \caption{Fixed-Len-2500/900 text-only on H200 (commit \texttt{ee8fdab1d}):
  Omni async-chunk route (default V1 deployment) versus thinker-only vLLM on one
  GPU. Medians over three repetitions.}
  \label{tab:omni-text}
  \begin{tabular}{l r r r}
    \toprule
    System & $C$ & TTFT mean (ms) & E2EL mean (s) \\
    \midrule
    Omni text-only & 1  & 75.0  & 3.79 \\
    Thinker vLLM   & 1  & 70.9  & 3.66 \\
    Omni text-only & 32 & 860.1 & 13.73 \\
    Thinker vLLM   & 32 & 562.4 & 13.67 \\
    \bottomrule
  \end{tabular}
\end{table}

\stitle{Streaming chunking}
Table~\ref{tab:omni-async} and the dashed curves in
Figure~\ref{fig:omni-fixedlen} toggle \texttt{--async-chunk} on Fixed-Len.
With chunking disabled, code2wav cannot start until the talker has finished
its codec sequence for a chunk boundary, which removes pipeline overlap
between thinker/talker production and audio egress.
The effect is modest at $C{=}1$ but grows with load: at $C{=}32$ the
no-async cell shows multi-second TTFT and mean E2EL well above the async
recipe, and mean RTF approaches one because the audio path spends more time
waiting behind upstream work rather than streaming incrementally.

\begin{table}[!htbp]
  \centering
  \small
  \caption{Async-chunk on versus off on Fixed-Len-2500/900 ($2{\times}$H100).}
  \label{tab:omni-async}
  \begin{tabular}{l r r r r}
    \toprule
    Mode & $C$ & TTFT mean (ms) & E2EL mean (s) & RTF mean \\
    \midrule
    async-chunk    & 1  & 96.4   & 18.51  & 0.117 \\
    no async-chunk & 1  & 115.8  & 25.07 & 0.169 \\
    async-chunk    & 8  & 271.9   & 31.91  & 0.236 \\
    no async-chunk & 8  & 446.7  & 37.62 & 0.247 \\
    async-chunk    & 32  & 507.8   & 72.63  & 0.630 \\
    no async-chunk & 32  & 3420.1  & 136.62 & 0.927 \\
    \bottomrule
  \end{tabular}
\end{table}

\begin{figure}[t]
  \centering
  \includegraphics[width=.9\linewidth]{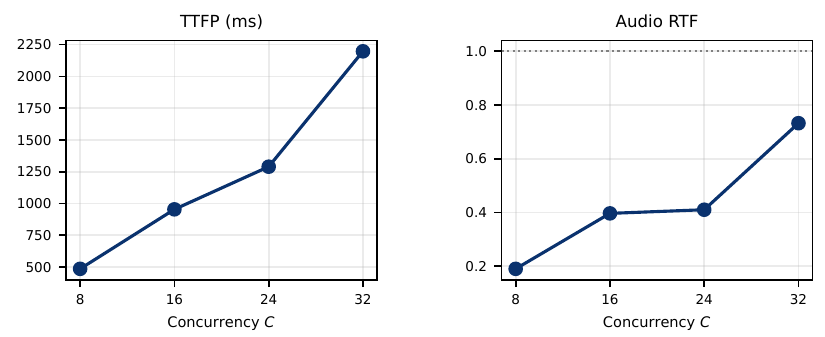}
  \caption{Qwen3-Omni 3-GPU replica=2 Fixed-Len sweep: mean audio TTFP and
  RTF versus concurrency.}
  \label{fig:omni-replica}
\end{figure}

\stitle{Replica scaling}
The three-GPU recipe duplicates talker and code2wav replicas to add downstream
capacity while the thinker layout stays on one device
(Table~\ref{tab:omni-replica}, Figure~\ref{fig:omni-replica}).
At $C{=}8$, lower mean audio TTFP and RTF versus the two-GPU async Fixed-Len
cell at the same concurrency indicate that extra audio replicas relieve
queueing in the mid-load regime where talker and code2wav were previously the
bottleneck.
At $C{=}32$, mean RTF is higher than on two GPUs despite replicas, which
suggests thinker admission and cross-stage handoff dominate before additional
code2wav copies help; TTFP and TPOT still climb as concurrent streams compete
for the same upstream stages.

\begin{table}[!htbp]
  \centering
  \small
  \caption{Talker/code2wav replica=2 on three H100s.}
  \label{tab:omni-replica}
  \begin{tabular}{r r r r r}
    \toprule
    $C$ & TTFT mean (ms) & TTFP mean (ms) & TPOT mean (ms) & RTF mean \\
    \midrule
    8  & 130.6 & 486.5  & 8.0  & 0.190 \\
    16  & 278.6 & 954.2  & 11.1  & 0.396 \\
    24  & 324.7 & 1289.7  & 16.2  & 0.410 \\
    32  & 911.6 & 2196.8  & 20.7  & 0.732 \\
    \bottomrule
  \end{tabular}
\end{table}

\subsection{TTS Serving}
\label{sec:eval-tts}

\begin{figure}[t]
  \centering
  \includegraphics[width=.96\linewidth]{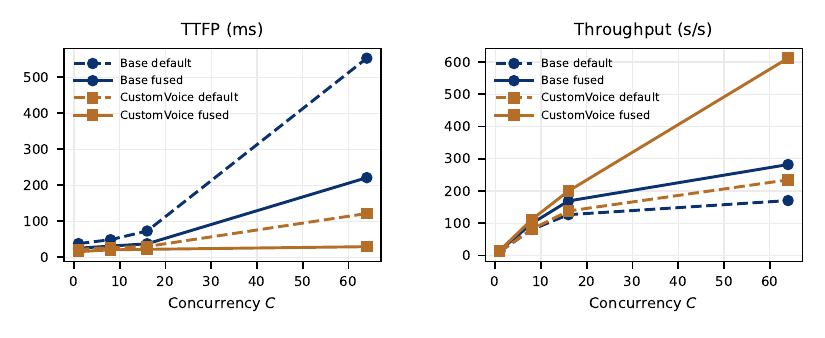}
  \caption{Qwen3-TTS median audio TTFP and audio throughput versus
  concurrency on H200, default two-stage (dashed) and fused single-stage
  (solid), for Base voice-clone and CustomVoice default-voice.}
  \label{fig:tts-scale}
\end{figure}

\stitle{TTS portfolio}
Table~\ref{tab:tts} and Figure~\ref{fig:tts-scale} evaluate Qwen3-TTS and Voxtral on Seed-TTS English texts.
Each Qwen3-TTS cell is measured under two deployments of the same model: the default two-stage recipe, which runs the talker and the Code2Wav decoder as separate stages on two GPUs, and the fused single-stage profile (\texttt{qwen3\_tts\_fused\_single\_gpu.yaml}), in which the talker uses a streaming codec decoder and returns PCM on one GPU.
Both deployments use MRv2.
The fused profile is faster in every cell while using half the GPUs.
On CustomVoice, it raises throughput from 81.1 to 111.6 audio-s/s at $C{=}8$ ($1.38\times$) and reaches 612.1 audio-s/s at $C{=}64$, while lowering median TTFP at $C{=}1$ from 21.5 to 15.3\,ms.
Base voice-clone achieves gains of $1.24\times$ at $C{=}8$ and $1.66\times$ at $C{=}64$.
Its TTFP is higher than CustomVoice because cloning pulls reference-speaker conditioning through the Base path.
The Base rows assume a server that has already processed the reference clips.
On a freshly started server, the per-clip caches (waveform, speaker embedding, reference codes) are empty, and median TTFP at $C{=}1$ is about 131\,ms for both deployments.
CosyVoice3's default deployment does not scale with load: throughput stays near 4.9 audio-s/s at every concurrency, so median TTFP grows to 51\,s at $C{=}64$.
The packed streaming profile batches Flow and HiFT decoding across requests and reaches 58.4 audio-s/s at $C{=}8$ and 78.4 at $C{=}64$ ($16\times$), with a median TTFP of 1.7\,s at $C{=}64$.
Whisper WER on 24 sampled outputs is 3.6\% versus 3.9\% for the default.
Under its default deployment, Higgs Audio v3 saturates near 29 audio-s/s from $C{=}16$ on, and median TTFP is 7.2\,s at $C{=}64$.
Its MRv2 profile keeps median TTFP between 80 and 150\,ms across the sweep and yields 311 audio-s/s at $C{=}64$ ($10.7\times$), with a Whisper WER of 4.9\% compared to 5.3\% for the default.
Fish Speech S2 Pro behaves the same at $C{=}1$ under both deployments, but the default saturates at 10.4 audio-s/s from $C{=}8$ on.
The high-concurrency profile reaches 40.4 audio-s/s at $C{=}64$ ($3.9\times$) and cuts median TTFP there from 22.3 to 2.2\,s, though its RTF still exceeds one at that load.
Whisper WER is 3.9\% versus 3.6\% for the default.
MOSS-TTS-Local ships three MRv2 profiles, all of which were measured.
The basic MRv2 profile is slower than the default (144 vs.\ 181 audio-s/s at $C{=}64$), while the high-concurrency and low-latency profiles both exceed it.
The low-latency profile reported in the table cuts median TTFP by 41--61\% across the sweep (285 to 145\,ms at $C{=}64$), raises throughput to 239 audio-s/s, and has a Whisper WER of 3.6\% versus 3.9\% for the default.
Voxtral-4B, served with its default two-GPU recipe, reaches 74.5\,ms median TTFP at $C{=}1$ and 43.4 audio-s/s at $C{=}8$.

\begin{table}[!htbp]
  \centering
  \small
  \caption{TTS models on Seed-TTS English texts (NVIDIA H200).
  ``Default'' is each model's default deployment; ``opt.''\ is its opt-in
  optimized profile: the fused single-stage Qwen3-TTS pipeline on one GPU
  (PR~\#8259; the default uses two GPUs) and CosyVoice3 packed streaming
  (PRs~\#8224 and \#8420), and the Higgs Audio v3 MRv2 H200 profile
  (PR~\#8226), and the Fish Speech high-concurrency single-GPU profile
  (PRs~\#3773 and \#8422), and the MOSS-TTS-Local MRv2 low-latency profile
  (PR~\#8213, without its optional MPS and multi-API-server launch settings);
  all four use one GPU in both deployments.
  Voxtral has no optimized profile.}
  \label{tab:tts}
  \begin{tabular}{l r rr rr rr}
    \toprule
    & & \multicolumn{2}{c}{TTFP med.\ (ms)} & \multicolumn{2}{c}{RTF med.} & \multicolumn{2}{c}{audio s/s} \\
    \cmidrule(lr){3-4}\cmidrule(lr){5-6}\cmidrule(lr){7-8}
    Model / task & $C$ & default & opt. & default & opt. & default & opt. \\
    \midrule
    Qwen3-TTS Base / clone      &  1 & 37.7 & 24.9 & 0.080 & 0.067 & 12.5 & 15.0 \\
    Qwen3-TTS Base / clone      &  8 & 48.6 & 30.8 & 0.095 & 0.077 & 80.5 & 99.5 \\
    Qwen3-TTS Base / clone      & 16 & 72.6 & 37.2 & 0.119 & 0.089 & 126.6 & 169.2 \\
    Qwen3-TTS Base / clone      & 64 & 552.8 & 220.9 & 0.347 & 0.198 & 170.4 & 282.1 \\
    Qwen3-TTS Custom / default  &  1 & 21.5 & 15.3 & 0.085 & 0.064 & 11.9 & 15.6 \\
    Qwen3-TTS Custom / default  &  8 & 26.6 & 20.3 & 0.095 & 0.068 & 81.1 & 111.6 \\
    Qwen3-TTS Custom / default  & 16 & 29.7 & 21.8 & 0.108 & 0.073 & 138.0 & 199.9 \\
    Qwen3-TTS Custom / default  & 64 & 121.3 & 29.1 & 0.247 & 0.100 & 234.6 & 612.1 \\
    Qwen3-TTS Custom / design   &  1 & 22.2 & 16.8 & 0.086 & 0.065 & 11.6 & 15.4 \\
    Qwen3-TTS Custom / design   &  8 & 29.9 & 21.6 & 0.102 & 0.071 & 74.4 & 104.7 \\
    CosyVoice3 / clone          &  1 & 445.9 & 325.8 & 0.250 & 0.067 & 4.0 & 14.8 \\
    CosyVoice3 / clone          &  8 & 3324.9 & 529.1 & 1.657 & 0.129 & 4.7 & 58.4 \\
    CosyVoice3 / clone          & 16 & 6974.5 & 746.9 & 3.265 & 0.205 & 4.8 & 77.1 \\
    CosyVoice3 / clone          & 64 & 51145.9 & 1673.8 & 11.450 & 0.608 & 4.9 & 78.4 \\
    Higgs Audio v3 / default    &  1 & 216.2 & 79.6 & 0.450 & 0.121 & 2.2 & 8.3 \\
    Higgs Audio v3 / default    &  8 & 246.0 & 83.4 & 0.496 & 0.121 & 15.6 & 61.6 \\
    Higgs Audio v3 / default    & 16 & 261.7 & 89.1 & 0.520 & 0.128 & 29.1 & 114.9 \\
    Higgs Audio v3 / default    & 64 & 7208.8 & 148.2 & 1.879 & 0.175 & 29.1 & 311.0 \\
    Fish Speech S2 Pro / clone  &  1 & 128.4 & 128.5 & 0.285 & 0.281 & 3.5 & 3.5 \\
    Fish Speech S2 Pro / clone  &  8 & 1695.1 & 199.4 & 0.771 & 0.465 & 10.3 & 16.6 \\
    Fish Speech S2 Pro / clone  & 16 & 4763.2 & 266.6 & 1.522 & 0.654 & 10.4 & 23.6 \\
    Fish Speech S2 Pro / clone  & 64 & 22326.2 & 2166.0 & 5.214 & 1.466 & 10.4 & 40.4 \\
    MOSS-TTS-Local / clone      &  1 & 47.0 & 27.6 & 0.087 & 0.075 & 11.5 & 13.4 \\
    MOSS-TTS-Local / clone      &  8 & 95.1 & 44.0 & 0.144 & 0.115 & 40.1 & 66.3 \\
    MOSS-TTS-Local / clone      & 16 & 140.5 & 55.1 & 0.176 & 0.145 & 88.1 & 105.7 \\
    MOSS-TTS-Local / clone      & 64 & 285.2 & 145.2 & 0.313 & 0.219 & 181.1 & 239.2 \\
    Voxtral-4B TTS              &  1 & 74.5 & --- & 0.107 & --- & 9.3 & --- \\
    Voxtral-4B TTS              &  8 & 117.3 & --- & 0.180 & --- & 43.4 & --- \\
    \bottomrule
  \end{tabular}
\end{table}

\FloatBarrier
\subsection{Image Generation}
\label{sec:eval-diffusion}

\stitle{Image generation}
Table~\ref{tab:qwen-image} covers Qwen-Image
CI still and edit recipes.
At $512^{2}$/20-step on one H100, mean latency near 2.4\,s is the single-stream
baseline; the step-execution sweep raises concurrency by batching denoising
work across requests, so per-request latency grows while req/s increases
because one GPU time-shares the same step loop.
Four-GPU Ulysses2+CFG2 with Cache-DiT cuts the $512^{2}$ cell and the
$1536^{2}$/35-step cell by reusing cached step computations and sharding
attention/CFG across devices, which shows up as lower latency and higher req/s
at $C{=}1$ without changing the exported peak-memory band by orders of magnitude.
Edit and layered rows in Table~\ref{tab:qwen-image-edit} repeat the same
Cache-DiT pattern on i2i workloads where conditioning tensors add work beyond
plain text-to-image.

\begin{table}[!htbp]
  \centering
  \small
  \caption{Qwen-Image still generation on H100 CI recipes.}
  \label{tab:qwen-image}
  \begin{tabular}{l r r r r}
    \toprule
    Config & $C$ & Latency (s) & req/s & Peak mem (MiB) \\
    \midrule
    $512^{2}$/20{,} 1$\times$H100 & 1 & 2.41 & 0.415 & 55{,}121 \\
    $512^{2}$/20{,} step-exec & 1 & 2.42 & 0.412 & 55{,}147 \\
    $512^{2}$/20{,} step-exec & 2 & 2.99 & 0.671 & 55{,}148 \\
    $512^{2}$/20{,} step-exec & 4 & 5.93 & 0.667 & 55{,}150 \\
    $512^{2}$/20{,} step-exec & 8 & 9.66 & 0.816 & 55{,}152 \\
    $512^{2}$/20{,} 4$\times$+Cache-DiT & 1 & 1.60 & 0.559 & 55{,}291 \\
    $1536^{2}$/35{,} 1$\times$H100 & 1 & 23.60 & 0.042 & 65{,}118 \\
    $1536^{2}$/35{,} Ulysses2+CFG2 & 1 & 8.22 & 0.121 & 54{,}501 \\
    $1536^{2}$/35{,} 4$\times$+Cache-DiT & 1 & 5.20 & 0.192 & 64{,}839 \\
    \bottomrule
  \end{tabular}
\end{table}
\begin{table}[!htbp]
  \centering
  \small
  \caption{Qwen-Image edit / layered single-device baselines (H100 recipes).}
  \label{tab:qwen-image-edit}
  \begin{tabular}{l r r r}
    \toprule
    Config & Latency (s) & req/s & Peak mem (MiB) \\
    \midrule
    Edit $512^{2}$/20 i2i & 8.84  & 0.113 & 58{,}572 \\
    Edit $512^{2}$/20 +Cache-DiT & 3.47  & 0.288 & 58{,}758 \\
    Edit-2511 $512^{2}$/20 (2 img) & 13.84  & 0.072 & 57{,}610 \\
    Edit-2511 $512^{2}$/20 +Cache-DiT & 3.59  & 0.278 & 58{,}075 \\
    Layered $640^{2}$/20 & 14.21  & 0.070 & 63{,}225 \\
    \bottomrule
  \end{tabular}
\end{table}

\FloatBarrier
\subsection{Video and World Models}
\label{sec:eval-world}

\stitle{Video generation}
Tables~\ref{tab:wan},~\ref{tab:cosmos}, and~\ref{tab:minimax} summarize Wan,
Cosmos3, and MiniMax-H3 video cells on the freeze-day recipes.
Wan2.2-I2V-A14B at $832{\times}480$/81 frames is compute-heavy on one H100;
USP2 with HSDP slicing parallelizes spatial and temporal shards so the same
schedule finishes faster (Table~\ref{tab:wan}), while the higher-resolution
$1280{\times}720$/121-frame USP2 point trades frame count and pixels for
much longer E2E time.
Cosmos3-Nano official-demo T2V, I2V, and V2V cluster near the same E2E latency
on two H100s because they share the demo schedule and step count, whereas T2I
is a separate shallow image head.
MiniMax-H3 T2V/TI2V/T2V+DLO stay in a tight band on four H100s; V2V is far
slower, which matches conditioning on an input video tensor in addition to text
rather than text-only generation.

\begin{table}[!htbp]
  \centering
  \small
  \caption{Wan2.2-I2V-A14B on H100.}
  \label{tab:wan}
  \begin{tabular}{l r r}
    \toprule
    Config & E2E mean (s) & req/s \\
    \midrule
    $832{\times}480$, 81f/4-step, 1$\times$ & 26.20 & 0.038 \\
    same, USP2+HSDP slicing                 & 18.04 & 0.055 \\
    $1280{\times}720$, 121f/4-step, USP2    & 96.19 & 0.010 \\
    \bottomrule
  \end{tabular}
\end{table}
\begin{table}[!htbp]
  \centering
  \small
  \caption{Cosmos3-Nano official-demo cells on $2{\times}$H100.}
  \label{tab:cosmos}
  \begin{tabular}{l r r r}
    \toprule
    Task & Schedule & E2E mean (s) & req/s \\
    \midrule
    T2I & $1024^{2}$/4-step              & 0.83  & 1.211 \\
    T2V & $1280{\times}720$, 189f/4-step & 28.88 & 0.035 \\
    I2V & $1280{\times}720$, 189f/4-step & 28.07 & 0.036 \\
    V2V & $1280{\times}720$, 189f/4-step & 28.17 & 0.036 \\
    \bottomrule
  \end{tabular}
\end{table}
\begin{table}[!htbp]
  \centering
  \small
  \caption{MiniMax-H3 on $4{\times}$H100
  ($1344{\times}768$, 209 frames, 8 steps).}
  \label{tab:minimax}
  \begin{tabular}{l r r}
    \toprule
    Task & E2E mean (s) & req/s \\
    \midrule
    T2V      & 31.15 & 0.032 \\
    T2V+DLO  & 30.75 & 0.033 \\
    TI2V     & 32.53 & 0.031 \\
    V2V      & 103.18 & 0.010 \\
    \bottomrule
  \end{tabular}
\end{table}

\FloatBarrier
\subsection{Full Duplex}
\label{sec:eval-duplex}

\stitle{Duplex and turn serving}
Table~\ref{tab:duplex} evaluates MiniCPM-o~4.5 on one H200 GPU using the default recipe (V1 runner, duplex session mode) and the protocol from the TTS table.
The challenge series increases concurrency on a fixed Seed-TTS workload, raising mean TTFP from 433\,ms at $C{=}1$ to 1533\,ms at $C{=}8$, mean RTF from 0.18 to 0.90, and throughput from 1.37 to 2.10 req/s.
Whisper WER across 128 outputs at $C{=}8$ is 1.6\%.
The opt-in turn-mode MRv2 profile (PR~\#8222) runs faster on this same workload, reaching a 658\,ms mean TTFP and 5.49 req/s at $C{=}8$.
However, it does not stop reliably, appending unrelated speech after the target sentence in 13\% of its outputs.
This behavior increases WER to 12.9\% (11.0\% using the native chat template), so it is not reported as a result.
The duplex Seed-TTS row at $C{=}1$ covers a single four-turn session on the realtime endpoint.
Because real-time input audio paces the session, its RTF stays near one, and because req/s counts sessions, this row serves as a distinct test fixture instead of a direct concurrency comparison.
These columns represent streaming health signals (TTFT/TTFP/RTF/throughput) rather than barge-in or admission-control protocol metrics.

\begin{table}[!htbp]
  \centering
  \small
  \caption{MiniCPM-o~4.5 on one NVIDIA H200 (commit \texttt{ee8fdab1d}), default
  recipe (V1 runner, duplex session mode). Same warm protocol as
  Table~\ref{tab:tts}; medians over three repetitions.}
  \label{tab:duplex}
  \begin{tabular}{l r r r r r}
    \toprule
    Cell & $C$ & TTFT mean (ms) & TTFP mean (ms) & RTF mean & req/s \\
    \midrule
    challenge & 1 & 96.5  & 433.3  & 0.180 & 1.368 \\
    challenge & 4 & 187.2 & 1028.9 & 0.487 & 1.949 \\
    challenge & 8 & 208.7 & 1533.4 & 0.895 & 2.100 \\
    duplex    & 1 & 341.7 & 341.7  & 1.088 & 0.059 \\
    \bottomrule
  \end{tabular}
\end{table}

\section{Related Work}
\label{sec:related}

\paragraph{LLM Serving Runtimes.}
Orca introduced iteration-level scheduling for autoregressive generation, and
PagedAttention coupled continuous batching with virtualized KV-cache
management~\cite{orca,vllm}.
SGLang extends this line with radix-prefix reuse and a runtime for structured
language-model programs~\cite{sglang}.
Sarathi-Serve uses chunked prefills to control decode interference, while
DistServe separates prefill and decode to optimize goodput under independent
latency objectives~\cite{sarathiserve,distserve}.
Mooncake and EPD further develop KV-centric transfer and
encoder--prefill--decode disaggregation~\cite{mooncake,epd}.
Enterprise stacks such as xLLM apply similar PD and EPD policies at the service
layer, with global KV management and accelerator-specific engine
optimizations~\cite{xllm}.
These systems optimize a single autoregressive decode loop or a multimodal
encoder--LLM path whose generated output remains a token stream.
They provide the scheduling, cache, and disaggregation substrate for
individual AR stages inside \sys{}, but they do not define execution semantics
across speech codecs, diffusion engines, and action generators in one request.

\paragraph{Diffusion and Multimodal Generation Serving.}
Diffusion Transformers and flow-matching generators motivate specialized
denoising runtimes for image and video synthesis~\cite{DiT}.
Diffusers supplies a broad model execution stack, while xDiT focuses on
parallel execution for large DiTs~\cite{diffusers,fang2024xdit}.
SGLang-Diffusion treats denoising as a first-class serving problem distinct
from LLM decode: it exposes denoising loops, CFG- and sequence-parallel
execution, cache backends, and recent AR+DiT disaggregation for hybrid image
generators~\cite{sglangdiffusion,GLMImage2026}.
Within a single diffusion stage, step-reduction caches such as TeaCache and
Cache-DiT reuse computation across denoising iterations~\cite{teacache,cache-dit}.
Multimodal LLM serving also disaggregates encoders from autoregressive
prefill and decode~\cite{epd,xllm}, but its product contract is still text
output conditioned on images, audio, or video.
Unified model families such as BAGEL interleave autoregressive understanding
with diffusion-style visual generation inside one transformer stack~\cite{bagel}.
Their serving needs therefore span both token decode and denoising, yet most
existing stacks optimize only one side.
\sys{} composes diffusion and AR engines as stages and assigns cross-stage
routing, streaming handoff, connector-backed transfer, and lifecycle
management to the surrounding runtime rather than to any single backend.

\paragraph{Multi-Stage Omni and Composite Serving.}
The closest line treats composite models as coordinated pipelines rather than
monolithic graphs.
vLLM-Omni introduced a stage graph with disaggregated AR, DiT, and other
engines connected by user-defined transfer functions~\cite{vllmomni}.
This report keeps that stage view but centers on an explicit orchestrator,
replica pools, streaming progress, stage-level KV ownership, and session
control on one control plane.
SGLang-Omni applies a similar stage-oriented design on SGLang schedulers, with
relay backends for tensor handoff and strong coverage of speech, ASR, and omni
chat~\cite{sglangomni}.
M* generalizes composition through Walk Graphs: requests execute as named walks
over a computation graph, with a Conductor scheduling workers under Sequential,
Parallel, Loop, and streaming-edge primitives~\cite{mstar2026}.
It reports gains over prior omni stacks on BAGEL and Qwen3-Omni configurations,
while optimizing graph-walk execution rather than the vLLM-integrated deploy
overlays, OpenAI-compatible endpoints, and session-backed serving emphasized
here.
Generic fabrics such as Ray Serve or Triton ensembles can connect arbitrary
stages, but they leave speech TTFP, async\_chunk overlap, stage KV transfer,
and duplex admission as application concerns.

\paragraph{Realtime, Robot, and Session-Oriented Serving.}
A second axis is interaction that outlives a single generate call.
Realtime duplex speech models such as MiniCPM-o expose listen-while-speak
semantics and turn-sensitive admission~\cite{minicpmo45,personaplex}.
World foundation models including Cosmos predict future observations or video
conditioned on past state~\cite{cosmos}.
Vision--language--action policies map perception and language to robot action
chunks~\cite{openvla,grootn1,dreamzero}.
In practice these paths are often hosted as specialized realtime servers,
model-specific demo scripts, or offline policy loops.
OpenPI, for example, provides a websocket policy server and thin client for
off-robot inference, but leaves batching, placement, and co-residency with
other modalities to the
operator~\cite{openpi}.
\sys{} instead treats lifecycle, state continuity, admission, and concurrent
I/O as shared session concerns on the multi-stage runtime, then specializes
the wire surface for duplex speech, conditional video, or OpenPI actions
without forking the orchestrator.

\section{Conclusion}
\label{sec:conclusion}

This report described the architecture of \sys{} as a stage-based serving
runtime with an explicit orchestrator, specialized engines, OmniConnector
transfer, intermediate-data management, and session control.
The evaluation draws those measurements from the public multimodal nightly CI
on NVIDIA H100, covering Qwen3-Omni speech serving, TTS, image and video
diffusion, Cosmos3/MiniMax video cells, and MiniCPM-o duplex health metrics.

Diffusion paged KV is now available on selected DiT pipelines under
scheduler-managed block allocation, but coverage and cache-aware admission
are still uneven across model families.
Broader session workloads, closed-loop robot evaluation, and measurements
across hardware platforms remain future work.

\section{Contributions}
\label{sec:contributions}

\paragraph{Core Contributors.}
\raggedright\small
Alex Brooks,
Nick Cao,
Weiqing Chen,
Didan Deng,
Yunyang Deng,
Han Gao,
Canlin Guo,
Yongxiang Huang,
Zeyu Huang,
Ziming Huang,
Minghui Jiang,
Shunyang Li,
Yueqian Lin,
Hongsheng Liu,
Weizhi Liu,
Ruixiang Ma,
Zifeng Mo,
Tun Jian Tan,
Cong Wang,
Roger Wang,
Yu Wang,
Zhipeng Wang,
Jared Wen,
Haiyan Wu,
Chendi Xue,
Peiqi Yin,
Yuanheng Zhao,
Chenguang Zheng,
Taichang Zhou,
Jiangyun Zhu.
\par

We thank all contributors to the \sys{}
project.\footnote{\url{https://github.com/vllm-project/vllm-omni/graphs/contributors}}

{
    \small
    \bibliographystyle{unsrt}
    \bibliography{refs}
}

\end{document}